\documentclass[a4paper,fleqn,usenatbib,useAMS]{mnras}

\usepackage{newtxtext,newtxmath}
\usepackage{graphicx}  
\usepackage{multicol}        
\usepackage{bm}    
\usepackage{pdflscape}  
\usepackage{multirow}
\usepackage{xcolor}
\usepackage{cuted}
\usepackage{hyperref}
\usepackage{booktabs}
\usepackage{subfigure}
\usepackage{academicons}
\usepackage{natbib}
\usepackage{orcidlink}
\usepackage{xspace}
\usepackage{amsmath}
\usepackage{tikz}
\usetikzlibrary{positioning,calc}

\newcommand{\orcid}[1]{\href{https://orcid.org/#1}{\textcolor[HTML]{A6CE39}{\aiOrcid}}}

\newcommand\dd{\mathrm d}

\newcommand\mg{\mathrm g}
\newcommand\kk{{\mathcal K_1-\mathcal K_2}}
\newcommand\mh{\mathrm h}
\newcommand\me{\mathrm e}
\newcommand\Msun{\mathrm M_\odot}

\DeclareRobustCommand{\VAN}[3]{#2}
\let\VANthebibliography\thebibliography
\def\thebibliography{\DeclareRobustCommand{\VAN}[3]{##3}\VANthebibliography}

\defcitealias{wanginflow1}{Paper~I}
\defcitealias{wanginflow3}{Paper~III}

\title[Inflow-driven galaxy evolution -- II]{Inflow-driven galaxy evolution -- II: A hierarchy in the scaling relations of star-forming galaxies}

\author[Wang]{Kai Wang,$^{1, 2}$\thanks{Contact e-mail: wkcosmology@gmail.com}\orcidlink{0000-0002-3775-0484}
  Carlton Baugh,$^{1,3}$\orcidlink{0000-0002-9935-9755}
  Sownak Bose,$^{1}$\orcidlink{0000-0002-0974-5266}
  Shaun Cole,$^{1}$\orcidlink{0000-0002-5954-7903}
  Carlos S. Frenk,$^{1}$\orcidlink{0000-0002-2338-716X}
  Cedric Lacey,$^{1}$\orcidlink{0000-0001-9016-5332}
  \newauthor
  Julio F. Navarro,$^{4, 5}$\orcidlink{0000-0003-3862-5076}
  Kyle A. Oman,$^{1,2}$\orcidlink{0000-0001-9857-7788}
  Amélie Saintonge,$^{6, 7}$\orcidlink{0000-0003-4357-3450}
  and Tom Theuns$^{1}$\orcidlink{0000-0002-3790-9520}
  \\
  $^{1}$Institute for Computational Cosmology, Department of Physics, Durham University, South Road, Durham, DH1 3LE, UK\\
  $^{2}$Centre for Extragalactic Astronomy, Department of Physics, Durham University, South Road, Durham, DH1 3LE, UK\\
  $^{3}$Institute for Data Science, Durham University, South Road, Durham, DH1 3LE, UK\\
  $^{4}$Department of Physics and Astronomy, University of Victoria, Victoria, BC, V8P 5C2, Canada\\
  $^{5}$Crimson Distinguished Professor, KU-KIST Green School, Korea University\\
  $^{6}$Max Planck Institute for Radio Astronomy, Auf dem Hügel 69, D-53121 Bonn, Germany\\
  $^{7}$Department of Physics \& Astronomy, University College London, London WC1E 6BT, UK
}

\date{Accepted XXX. Received YYY; in original form ZZZ}

\pubyear{\the\year{}}

\begin{document}
\label{firstpage}
\pagerange{\pageref{firstpage}--\pageref{lastpage}}
\maketitle

\begin{abstract}
  The star-forming main sequence (SFMS; $M_\star$--$\rm sSFR$), the gas-to-stellar mass relation ($M_\star$--$M_\mg/M_\star$), and the mass--metallicity relation (MZR; $M_\star$--$Z_\mg$) are three core scaling relations for star-forming galaxies, and together they encode the information about the baryon cycle that operates during galaxy evolution.
  We investigate how these relations emerge by modelling the gas flows that regulate the baryon cycle, using both an ideal and a cosmological gas flow model, and identify a hierarchy among the three relations, with one further parameter entering the scatter budget at each step.
  In our model, the SFMS and its scatter are set primarily by the formation time of the host halo, with a weak dependence on the star formation efficiency and the mass-loading factor.
  The gas-to-stellar mass relation follows from the ratio of the sSFR to the star formation efficiency, and inherits the scatter of both.
  The MZR in turn follows from the gas-to-stellar mass relation together with the mass-loading factor, their relative contributions depending on the galaxy evolutionary state between the inflow-driven regime and equilibrium.
  This hierarchy naturally introduces further correlations among the three relations, such as the fundamental plane of star formation ($M_\star$--$M_\mg$--$\mathrm{SFR}$), the fundamental metallicity relation ($M_\star$--$\mathrm{SFR}$--$Z_\mg$) and its gaseous counterpart ($M_\star$--$M_\mg$--$Z_\mg$), and the correlations of each relation with galaxy size, which traces the star formation efficiency through the star formation law.
  Together, the three core scaling relations and the higher-dimensional correlations among them follow from a single framework, providing a foundation for linking the observables of star-forming galaxies to the underlying baryon cycle.
\end{abstract}




\section{Introduction}
\label{sec:introduction}

Galaxies form within dark matter haloes and are supplied with baryons that are accreted as the halo grows \citep{whiteCoreCondensationHeavy1978}.
The accreted gas can join the interstellar medium (ISM) by two routes.
Part of the accreted gas is shock-heated to the virial temperature and must radiate this energy away before it can reach the galaxy. Part reaches the ISM directly through cold-mode accretion \citep[e.g.][]{keresHowGalaxiesGet2005, dekelColdStreamsEarly2009}.
The dense gas that reaches the ISM forms stars at an efficiency that depends on the density and thermal state of the gas.
Observationally, the surface density of the star formation rate correlates with the gas surface density \citep{kennicuttStarFormationGalaxies1998} and the stellar surface density \citep{shiExtendedSchmidtLaw2011}.
Each generation of star formation returns metals to the ISM together with a fraction of the gas from which it formed, and supernova explosions drive metal-enriched ISM out of the galaxy and, in sufficiently shallow potentials, out of the halo.
This baryon cycle constitutes the basic picture of how cosmic gas fuels star formation in a galaxy \citep[e.g.][]{coleHierarchicalGalaxyFormation2000, mitchellGalacticInflowWind2020, mitchellGalacticOutflowRates2020, maFountainPatternBaryon2026a}.

Over the past two decades galaxy surveys have established three tight scaling relations among stellar mass, star formation rate (SFR), gas mass, and gas metallicity.
First, star-forming galaxies occupy a narrow locus in the plane of stellar mass ($M_\star$) and specific star formation rate (${\rm sSFR}\equiv{\rm SFR}/M_\star$), known as the star-forming main sequence \citep[SFMS;][]{brinchmannPhysicalPropertiesStarforming2004, noeskeStarFormationAEGIS2007, salimUVStarFormation2007}.
The sSFR is nearly independent of stellar mass below $M_\star \approx 10^{10.5}\,\Msun$, with its normalisation decreasing towards higher stellar masses and increasing with redshift \citep{daddiMultiwavelengthStudyMassive2007, elbazReversalStarFormationdensity2007, pengMassEnvironmentDrivers2010, whitakerStarFormationMass2012, speagleHighlyConsistentFramework2014, renziniObjectiveDefinitionMain2015, popessoMainSequenceStarforming2023, huangExploringIntrinsicScatter2023}.
Second, the gas mass\footnote{Here the gas mass is the total atomic and molecular gas mass of the ISM. Observationally, we primarily focus on atomic gas, which dominates the reservoir of star-forming galaxies \citep{saintongeColdInterstellarMedium2022a, chowdhuryAtomicGasDominates2022}.} lies on a power-law relation with stellar mass for star-forming galaxies \citep{haynesNeutralHydrogenIsolated1984, catinellaGALEXAreciboSDSS2010, tacconiHighMolecularGas2010, tacconiPHIBSSUnifiedScaling2018, catinellaXGASSTotalCold2018, tacconiEvolutionStarFormingInterstellar2020, saintongeColdInterstellarMedium2022a}.
This relation has a slope of $\approx 0.4$ in the range  $10^{9} < M_\star/\Msun < 10^{11}$ at $z\approx 0$ \citep{saintongeColdInterstellarMedium2022a}.
Because the slope is well below unity, low-mass galaxies have higher gas fractions than massive galaxies.
Measuring the atomic gas content of individual galaxies becomes difficult beyond $z \sim 0.1$, and spectral stacking is required to extend the measurements to $z \approx 1$.
The available constraints indicate that the atomic gas fraction changes little out to $z \approx 0.4$ at fixed stellar mass \citep{zhangWeakEvolutionCosmic2026} and rises by a factor of $\approx 3.5$ by $z \approx 1$ \citep{chowdhuryAtomicGasDominates2022, chowdhuryAtomicGasScaling2022}.
Finally, the gas-phase metallicity, measured as the abundance of oxygen relative to hydrogen in star-forming gas, follows a scaling relation with stellar mass known as the mass--metallicity relation \citep[MZR;][]{tremontiOriginMassMetallicityRelation2004, curtiMassmetallicityFundamentalMetallicity2020, scholteAtomicGasSequence2024, lyuDominantRoleCoplanar2025}.
The relation has a power-law slope of $\approx 0.3$ at the low-mass end and flattens at the massive end, and its normalisation decreases with redshift \citep{erbMassMetallicityRelation2006, maiolinoAMAZEEvolutionMassmetallicity2008, sandersMOSDEFSurveyEvolution2021, liMassMetallicityRelationDwarf2023, heEarlyResultsGLASSJWST2024, jainUniformAnalysisGasphase2026, liInsightsMetalEnrichment2025, lamJADESMassMetallicityFundamental2026}.

These three scaling relations are not independent in either observations or galaxy formation models, and the correlations among them are captured by a set of higher-dimensional relations.
Galaxies occupy a two-dimensional surface in the space of stellar mass, SFR, and gas-phase metallicity, known as the fundamental metallicity relation \citep[FMR;][]{ellisonCluesOriginMassMetallicity2008, mannucciFundamentalRelationMass2010, andrewsMassMetallicityRelationDirect2013, dayalPhysicsFundamentalMetallicity2013, salimCriticalLookMassMetallicityStar2014a, curtiMassmetallicityFundamentalMetallicity2020}, and this surface is nearly independent of redshift out to $z \approx 3$ \citep{huntCoevolutionMetallicityStar2016, sandersMOSDEFSurveyEvolution2021, curtiJADESInsightsLowmass2024}.
Replacing the SFR with the atomic gas mass yields the gaseous FMR, and the relation is consistently tighter with gas mass than with SFR as the third parameter \citep{bothwellFundamentalRelationMetallicity2013, jimmyGasPhaseMass2015, bothwellGalaxyMetallicitiesDepend2016, boardmanFundamentalMetallicityRelation2026}.
Galaxies likewise occupy a two-dimensional surface in the space of stellar mass, SFR, and gas mass out to $z \sim 3$ \citep{santiniEvolutionDustGas2014, douHaloesGalaxiesII2021}, which is also recovered in the EAGLE simulation \citep{schayeEAGLEProjectSimulating2015} and termed the `fundamental plane of star formation' \citep{lagosFundamentalPlaneStar2016}.

These higher-dimensional relations manifest as scatter in the three primary relations, so studying that scatter is one way to uncover how the primary relations connect.
First, using the EAGLE simulation, \citet{mattheeOriginScatterStar2019} found that the scatter of the SFMS correlates with the formation time of the host dark matter halo, with galaxies below the SFMS residing preferentially in early-formed haloes \citep{boucheImpactColdGas2010, duttonOriginGalaxyStarformationrate2010, blankNIHAOXXVINature2021, chenMAHGICModelAdapter2021, chenHowEmpiricallyModel2021a}.
An alternative investigation attributes the scatter to stochasticity driven by bursty star formation and feedback \citep[e.g.][]{faucher-giguereModelOriginBursty2018, caplarStochasticModellingStarformation2019, wangElevationSuppressionStar2019, tacchellaStochasticModellingStarformation2020, wangVariabilityStarFormation2020, wangVariabilityStarFormation2020a, guiEpisodicStarFormation2026}.
These studies model the star formation history as a median trend plus a stochastic component.
They only account for the scatter about the trend, and so are complementary to a theory of the SFMS rather than being a complete theory of the SFMS themselves.
Observational studies commonly decompose the SFR into the product of gas mass and star formation efficiency, then attribute the scatter of the SFMS to variation in one or both factors \citep[e.g.][]{linALMaQUESTSurveyMolecular2019, feldmannLinkStarFormation2020, saintongeColdInterstellarMedium2022a}.
This leaves the origin of the gas mass itself unexplained, so the decomposition relocates the question of the SFMS origin instead of answering it.
Second, at fixed stellar mass, the gas mass of star-forming galaxies correlates positively with galaxy size and with the SFR \citep{zhangEstimatingHIGas2009, catinellaXGASSTotalCold2018, saintongeColdInterstellarMedium2022a}.
Third, the gas-phase metallicity correlates with several galaxy properties at fixed stellar mass.
It correlates negatively with the SFR \citep[e.g.][]{ellisonCluesOriginMassMetallicity2008, mannucciFundamentalRelationMass2010, curtiMassmetallicityFundamentalMetallicity2020}, with galaxy size \citep[e.g.][]{ellisonCluesOriginMassMetallicity2008, wangConnectingCompactStarforming2018, sanchez-menguianoMoreFundamentalFundamental2024, maRevisitingFundamentalMetallicity2024, boardmanCompetingEffectsRecent2025, boardmanFundamentalMetallicityRelation2026, wangOriginGalaxySizestellar2026}, and with the gas mass \citep[e.g.][]{bothwellFundamentalRelationMetallicity2013, jimmyGasPhaseMass2015}.

These scaling relations have been studied extensively in observations and in galaxy formation models \citep[e.g.][]{derossiGalaxyMetallicityScaling2017, mattheeOriginScatterStar2019, donnariStarFormationActivity2019, garciaDoesFundamentalMetallicity2024, garciaDoesFundamentalMetallicity2025, shardaEvolutionGalaxyGasphase2026a}, but they are still largely treated one at a time, and a unified picture of how the three relate to one another is still lacking.
Many models also remain in tension with observations.
Several semi-analytic models produce an MZR that barely evolves with redshift \citep[e.g.][]{luSemianalyticModelsCANDELS2014, somervilleStarFormationSemianalytic2015, guoGalaxiesEAGLEHydrodynamical2016, chenTwophaseModelGalaxy2025}, and hydrodynamical simulations differ substantially in their predicted evolution of the FMR \citep{garciaDoesFundamentalMetallicity2024, garciaDoesFundamentalMetallicity2025}.

Part of the difficulty is that current semi-analytic models and hydrodynamical simulations have grown sufficiently sophisticated that the dependence of a given observable on a given physical process is no longer transparent and readily isolated.
Those models are therefore used mainly for comparison with observational data; an agreement does not always identify the driving mechanism by itself \citep[e.g.][]{derossiGalaxyMetallicityScaling2017, donnariStarFormationActivity2019, shardaEvolutionGalaxyGasphase2026a}, and a disagreement cannot easily be traced to the failure of a particular process \citep[e.g.][]{garciaDoesFundamentalMetallicity2024, garciaDoesFundamentalMetallicity2025}.
Here we adopt a much simpler model\footnote{Our model is built on the same mass-continuity equations as earlier gas-regulator models \citep{daveAnalyticModelEvolution2012, lillyGASREGULATIONGALAXIES2013}, but these models assume that the gas reservoir is in (quasi-)equilibrium. They therefore apply only to massive galaxies at low redshift, which lie close to equilibrium and for which the assumption is justified. \S\,5.1 of \citetalias{wanginflow1} discusses this limitation in detail.} to reveal the basic mechanism through which these relations emerge.
It trades realism for exactly this transparency: every observable can be written in closed form in terms of the processes that set it, and the dependence on each can be read off directly.
Building on the framework of \citet[][hereafter \citetalias{wanginflow1}]{wanginflow1}, we start from the two mass-continuity equations that govern the baryon cycle and establish a hierarchy among the three scaling relations.

This paper is organised as follows.
\S\,\ref{sec:perspective_ideal} builds intuition from the ideal analytic model, in which the gas inflow rate, the star formation efficiency, and the mass-loading factor are held constant.
\S\,\ref{sec:perspective_cosmological} performs the numerical tests in the minimal cosmological gas flow model of \citetalias{wanginflow1} and describes the hierarchy among the three relations.
\S\,\ref{sec:what_shape} shows how the mass and redshift dependences of the three relations are shaped.
\S\,\ref{sec:discussion} discusses the origin of each relation in the light of observations and of competing scenarios, and the role of halo mass.
\S\,\ref{sec:summary} summarises our conclusions.

\section{An analytic perspective from the ideal gas flow model}
\label{sec:perspective_ideal}

We start from the gas flow model of \citetalias{wanginflow1}, which we briefly review here \citep[see also][]{coleHierarchicalGalaxyFormation2000}.
Gas is supplied at a rate $\Phi$, converted into stars at a rate ${\rm SFR} = \epsilon M_\mg$ set by the star formation efficiency $\epsilon$, and expelled at a rate $\eta\times {\rm SFR}$ set by the mass-loading factor $\eta$.
We take the inflowing gas to be metal-free, neglecting the re-accretion of previously ejected, metal-enriched gas \citep[e.g.][]{mitchellGalacticInflowWind2020}.
A fraction $R$ of the mass formed into stars is returned to the ISM under the instantaneous recycling approximation, and metals are released at a rate $y\times {\rm SFR}$, where $y$ is the yield.
Both $R$ and $y$ are held constant, as set by the IMF, metallicity, and stellar evolution (Table~\ref{tab:parameters}).
The approximation is well justified for oxygen, the most abundant metal by mass \citep[e.g.][]{asplundChemicalCompositionSun2009} and the element through which the gas metallicity is measured \citep[e.g.][]{maiolinoReMetallicaCosmic2019}, because it is released mainly by core-collapse supernovae from progenitors with lifetimes of $\la 40$~Myr \citep[e.g.][]{nomotoNucleosynthesisStarsChemical2013}, much shorter than the timescales of galaxy evolution considered here, such as the equilibrium timescale, $\tau_{\rm eq} \approx 3.4$~Gyr at $M_\star = 10^{10}\,\Msun$ and $z = 0$ (equation~\ref{eq:tau_eq}).
We also neglect mergers, which contribute substantially to stellar mass growth only above $M_\star \approx 10^{10.5}\,\Msun$ \citep[e.g.][]{davisonEAGLEsViewEx2020, proctorWeakConnectionStellar2025, liPhysicalProcessesCoevolution2025,
  wangForgedQuenchingMorphological2026, wangRuffledFeathersMergerdriven2026, chenTwophaseModelGalaxy2026}.
The gas and metal mass-continuity equations then follow, in the form given by \citet{coleHierarchicalGalaxyFormation2000}\footnote{Note that our yield $y$ is denoted $p$ in \citet{coleHierarchicalGalaxyFormation2000}.},
\begin{align}
  \frac{\dd M_\mg}{\dd t}         & = \Phi - (1 - R + \eta)\,\epsilon\,M_\mg,
  \label{eq:cont_gas}                                                         \\[4pt]
  \frac{\dd (Z_\mg M_\mg)}{\dd t} & = y\,\epsilon\,M_\mg
  - (1 - R + \eta)\,\epsilon\,M_\mg\,Z_\mg.
  \label{eq:cont_metal}
\end{align}
The stellar mass then follows as the time integral of the SFR reduced by the returned fraction, $M_\star = \int (1-R)\,\epsilon\,M_\mg\,\dd t$.

Solving these two evolution equations requires the gas inflow rate $\Phi$, the star formation efficiency $\epsilon$, and the mass-loading factor $\eta$.
We first consider the ideal case of \citetalias{wanginflow1}, in which all three variables are constants.
In this simple case, the coupled equations can be solved analytically,
\begin{align}
  M_\mg(t)   & = \Phi\,\tau_{\rm eq} \left(1 - \me^{-t/\tau_{\rm eq}}\right),
  \label{eq:mgas_ana}                                                         \\[4pt]
  Z_\mg(t)   & = y\,\epsilon\left(\tau_{\rm eq}
  - \frac{t}{\me^{t/\tau_{\rm eq}} - 1}\right),
  \label{eq:zgas_ana}                                                         \\[4pt]
  M_\star(t) & = (1-R)\,\epsilon\,\tau_{\rm eq}\,\Phi
  \left[t - \tau_{\rm eq} \left(1 - \me^{-t/\tau_{\rm eq}}\right)\right],
  \label{eq:mstar_ana}
\end{align}
where $t$, the \textit{galaxy evolution time}, is the time elapsed since the onset of gas accretion.
When $t \gg \tau_{\rm eq}$, the gas mass, the gas metallicity, and the SFR all become constant, and we refer to this limit as the ``equilibrium regime''.
For $t \ll \tau_{\rm eq}$, the gas mass, the metallicity, and the SFR instead increase with time, and we refer to this stage as the ``inflow-driven regime'' \citep[see][for details]{wanginflow1}.
The equilibrium timescale is defined as
\begin{equation}
  \tau_{\rm eq} = \frac{1}{(1-R+\eta)\,\epsilon},
  \label{eq:tau_eq}
\end{equation}
which determines the timescale required to approach equilibrium.
The SFR of galaxies follows directly as ${\rm SFR}=\epsilon M_\mg$.

Within this model, the gas mass, stellar mass, SFR and gas metallicity are determined by the three parameters $\Phi$, $\epsilon$, and $\eta$, together with the evolution time $t$.
The inflow rate enters linearly into every extensive quantity: the gas mass, the stellar mass, the SFR, and the metal mass in gas $Z_\mg M_\mg$.
Ratios of these quantities are therefore independent of $\Phi$.
This applies to the sSFR, the gas fraction ($M_\mg/M_\star$), and the gas metallicity ($Z_\mg$), which is itself the ratio of metal mass in gas to gas mass.
Three degrees of freedom therefore remain, $\epsilon$, $\eta$, and $t$, and we now ask how each of them propagates into the three observables.

We first consider the sSFR.
Dividing equation~\eqref{eq:mgas_ana} by equation~\eqref{eq:mstar_ana} and using ${\rm sSFR} = \epsilon M_\mg/M_\star$ gives
\begin{equation}
  {\rm sSFR} = \frac{1}{(1-R)\,\mathcal K_2^\prime(t/\tau_{\rm eq})}\,\frac{1}{t},
  \qquad
  \mathcal K_2^\prime(x) = \frac{x - 1 + \me^{-x}}{x\left(1 - \me^{-x}\right)},
  \label{eq:ssfr}
\end{equation}
where $\mathcal K_2^\prime(x)$ increases monotonically from $1/2$ for $x \ll 1$ to unity for $x \gg 1$.
The star formation efficiency and the mass-loading factor enter only through $\mathcal K_2^\prime (t/\tau_{\rm eq})$, so at fixed $t$ they change the sSFR by no more than a factor of two.
The sSFR is primarily set by the galaxy evolution time, $t$, with galaxies at larger $t$ having lower sSFR.
Galaxies in early-formed haloes began accreting earlier and so have larger $t$ at fixed stellar mass and given cosmic time.
This offers a natural explanation for the result of \citet{mattheeOriginScatterStar2019}, who found that galaxies above the SFMS are hosted by late-formed haloes and galaxies below it by early-formed haloes \citep[see also][]{duttonOriginGalaxyStarformationrate2010, blankNIHAOXXVINature2021}.

We next consider the gas fraction ($M_\mg/M_\star$), which follows from the identity
\begin{equation}
  \frac{M_\mg}{M_\star} = \frac{\rm sSFR}{\epsilon} = \frac{1}{(1-R)\,\mathcal K_2^\prime(t/\tau_{\rm eq})}\,\frac{1}{t}\times \frac{1}{\epsilon}.
\end{equation}
The gas fraction inherits the evolution-time ($t$) dependence of the sSFR and scales inversely with the star formation efficiency.
The mass-loading factor enters only through the dependence of the sSFR on $\mathcal K_2^\prime$, which varies by less than a factor of two, as detailed above.

We finally consider the gas metallicity ($Z_\mg$).
The analytic solution, equations~\eqref{eq:mgas_ana} to \eqref{eq:mstar_ana}, can be recast to express the gas metallicity in terms of the gas fraction and the mass-loading factor, which we refer to as the $\kk$ framework (see \citetalias{wanginflow1} for the detailed derivation),
\begin{align}
   & Z_\mg = \frac{y}{1 - R} \left(\frac{M_\mg}{M_\star}\right)^{-1}
  \mathcal{K}_1\left(\frac{t}{\tau_{\rm eq}}\right), \quad
  \mathcal{K}_1(x) = \frac{1 - \me^{-x} - x\,\me^{-x}}{x - 1 + \me^{-x}},
  \label{eq:master}                                                     \\[4pt]
   & \frac{1 - R + \eta}{1 - R} \left(\frac{M_\mg}{M_\star}\right)^{-1}
  = \mathcal{K}_2\left(\frac{t}{\tau_{\rm eq}}\right), \quad
  \mathcal{K}_2(x) = \frac{x - 1 + \me^{-x}}{1 - \me^{-x}}.
  \label{eq:t2tau}
\end{align}
Equations~\eqref{eq:master} and~\eqref{eq:t2tau} together determine the gas metallicity from the gas fraction and the mass-loading factor.
All three parameters therefore shape the metallicity: the galaxy evolution time $t$ and the star formation efficiency $\epsilon$ enter through the gas fraction, and the mass-loading factor $\eta$ acts directly.

We refer the reader to \citetalias{wanginflow1} for a detailed discussion of equations~\eqref{eq:master} and \eqref{eq:t2tau}, and briefly review their behaviour here for convenience.
In the inflow-driven regime, $t/\tau_{\rm eq} \ll 1$, we have $\mathcal K_1 \rightarrow 1$ and $\mathcal K_2 \rightarrow x/2$, so equation~\eqref{eq:master} reduces to $Z_\mg = [y/(1-R)]\,(M_\mg/M_\star)^{-1}$ and the metallicity is set by the gas fraction alone.
In equilibrium, $t/\tau_{\rm eq} \gg 1$, we have $\mathcal K_1 \rightarrow 1/x$ and $\mathcal K_2 \rightarrow x$, and the factor $x$ cancels between the two equations to give $Z_\mg = y/(1-R+\eta)$, so the gas fraction drops out and the metallicity is set by the mass-loading factor alone.
Between these limits the metallicity is set by both the gas fraction and the mass-loading factor, with their relative weight determined by the evolution stage $t/\tau_{\rm eq}$.

The galaxy properties $M_\star$, $M_\mg$, and $Z_\mg$, and the scaling relations among them, therefore form a hierarchy.
The sSFR is primarily set by the galaxy evolution time, with a weak modulation from the star formation efficiency and the mass-loading factor.
The gas fraction follows through the identity $M_\mg/M_\star = {\rm sSFR} / \epsilon$, which introduces an explicit dependence on the star formation efficiency.
The gas metallicity then follows from the gas fraction and the mass-loading factor through the $\mathcal K_1$--$\mathcal K_2$ framework.
At each step one additional parameter enters: $t$ sets the sSFR, $\epsilon$ enters at the gas fraction, and $\eta$ enters at the gas metallicity.
We also note that none of the specific quantities, the sSFR, $M_\mg/M_\star$, and $Z_\mg$, depends on the inflow rate $\Phi$.

The hierarchy holds in both the inflow-driven and equilibrium regimes, though two of its features depend on the evolutionary stage.
First, the sSFR changes by no more than a factor of two across the transition, since $\mathcal K_2^\prime$ varies only between $1/2$ and 1.
Second, the quantity that sets the gas metallicity changes entirely between the two regimes.
In the inflow-driven regime $\mathcal K_1 \rightarrow 1$ and $Z_\mg = [y/(1-R)]\,(M_\mg/M_\star)^{-1}$, so the gas metallicity is determined by the gas fraction.
In equilibrium $Z_\mg = y/(1-R+\eta)$, so the gas metallicity is determined by the mass-loading factor alone.

Although the roles of the star formation efficiency and mass-loading factor differ between the two limiting regimes, changes in either affect the SFR and stellar mass proportionally and hence leave the sSFR almost unchanged \citep[see also][]{boucheImpactColdGas2010, duttonOriginGalaxyStarformationrate2010, lillyGASREGULATIONGALAXIES2013, pengHaloesGalaxiesDynamics2014}.
In the inflow-driven regime ${\rm SFR} \rightarrow \epsilon\,\Phi\,t$ and $M_\star \rightarrow (1-R)\,\epsilon\,\Phi\,t^2/2$ are both independent of $\eta$ and both proportional to $\epsilon$, so a change in $\epsilon$ scales $M_\star$ and ${\rm SFR}$ proportionally and leaves the sSFR unchanged.
In equilibrium ${\rm SFR} = \Phi/(1-R+\eta)$ and $M_\star = (1-R)\,\Phi\,t/(1-R+\eta)$ are both independent of $\epsilon$ and both suppressed by the same factor $1/(1-R+\eta)$, so a change in $\eta$ also leaves the sSFR unchanged.
Their effect on the sSFR is therefore confined to altering the transition from the inflow-driven regime to equilibrium.

The mass-loading factor cancels in the same way from the gas fraction.
The mass-loading factor does not appear in the gas fraction in the inflow-driven regime, and in equilibrium it suppresses the gas mass, $M_\mg = \Phi/[(1-R+\eta)\,\epsilon]$, and the stellar mass by the same factor $1/(1-R+\eta)$, which cancels in the ratio.
As with the sSFR, the mass-loading factor therefore acts on the gas fraction only through the transition between the two regimes.

In summary, we have shown how the three parameters, the galaxy evolution time, the star formation efficiency, and the mass-loading factor, shape the evolution of a galaxy, and how the hierarchy follows from mass and metal conservation alone.
In \S\,\ref{sec:perspective_cosmological} we adopt specific forms for the three parameters and examine how the resulting galaxy population evolves.

\section{A semi-analytic perspective from the cosmological gas flow model}
\label{sec:perspective_cosmological}

\subsection{The cosmological gas flow model}
\label{sub:The cosmological gas flow model} 

\begin{table}
  \centering
  \caption{
    Parameters of the cosmological gas flow model.
    Anchoring masses are $M_\lambda = 10^{12}\, \Msun$ and $M_\epsilon = M_\eta = 10^{10}\,\Msun$.}
  \label{tab:parameters}
  \begin{tabular}{lll}
    \hline
    Symbol          & Parameter                       & Best-fit value                                                                                                          \\
    \hline
    \multicolumn{3}{l}{\textbf{Cooling efficiency}: $\lambda=\frac{\lambda_0 (1 + z)^{\lambda_z}}{(M_\mh / M_\lambda)^{\lambda_{m_1}}  + (M_\mh / M_\lambda)^{\lambda_{m_2}}}$} \\[10pt]
    $\lambda_{0}$   & Cooling normalisation           & $0.35$                                                                                                                  \\
    $\lambda_{m_1}$ & Low-mass suppression slope      & $-0.60$                                                                                                                 \\
    $\lambda_{m_2}$ & High-mass suppression slope     & $0.20$                                                                                                                  \\
    $\lambda_{z}$   & Cooling redshift exponent       & $0.20$                                                                                                                  \\[8pt]
    \hline
    \multicolumn{3}{l}{\textbf{Star formation efficiency (SFE)}: $\epsilon=\epsilon_0\left(\frac{M_\star}{M_\epsilon}\right)^{\epsilon_m} (1 + z)^{\epsilon_z}$}                \\[10pt]
    $\epsilon_0$    & SFE normalisation               & $0.38\;\mathrm{Gyr}^{-1}$                                                                                               \\
    $\epsilon_m$    & SFE stellar mass slope          & $0.33$                                                                                                                  \\
    $\epsilon_z$    & SFE redshift exponent           & $1.1$                                                                                                                   \\[8pt]
    \hline
    \multicolumn{3}{l}{\textbf{Mass-loading}: $\eta = \eta_0\left(\frac{M_\star}{M_\eta}\right)^{\eta_m}$}                                                                      \\[10pt]
    $\eta_0$        & Mass-loading normalisation      & $0.22$                                                                                                                  \\
    $\eta_m$        & Mass-loading stellar mass slope & $-0.30$                                                                                                                 \\[4pt]
    \hline
    \multicolumn{3}{l}{\textbf{IMF and yield}}                                                                                                                                  \\[2pt]
    $R$             & Mass return fraction            & $0.44$                                                                                                                  \\
    $y$             & Metal yield                     & $0.012$                                                                                                                 \\
    \hline
  \end{tabular}
\end{table}

The ideal model of \S\,\ref{sec:perspective_ideal} holds constant the inflow rate ($\Phi$), the star formation efficiency ($\epsilon$), and the mass-loading factor ($\eta$).
We now use the cosmological gas flow model of \citetalias{wanginflow1} to revisit how these quantities affect the three scaling relations.
We summarise the components here and refer the reader to \citetalias{wanginflow1} for the modelling and calibration details.
In this model these quantities instead evolve with halo mass and redshift.

The model integrates the two mass-continuity equations, equations~\eqref{eq:cont_gas} and \eqref{eq:cont_metal}, with a prescription for each of the three quantities they depend on: the gas inflow rate $\Phi$, the star formation efficiency $\epsilon$, and the mass-loading factor $\eta$.

The gas inflow rate is taken to be the product of the baryonic accretion rate at the halo scale and a cooling efficiency $\lambda$,
\begin{equation}
  \Phi = \lambda(M_\mh, z)\, f_{\rm b}\, \dot M_\mh ,
\end{equation}
where $f_{\rm b} = \Omega_{\rm b}/\Omega_{\rm m}$ is the universal baryon fraction.
We take $\lambda$ to be the fraction of the accreted baryons that reaches the ISM, so that it encapsulates several distinct processes: the radiative cooling of the halo gas, which becomes inefficient in massive haloes \citep{sutherlandCoolingFunctionsLowDensity1993, moTwophaseModelGalaxy2024}; preventive feedback, which suppresses accretion in low-mass haloes \citep{daveAnalyticModelEvolution2012, mitraEquilibriumModelConstraints2015, luImportancePreventiveFeedback2017, yuStellarFeedbackDrives2026}; AGN feedback \citep{bowerBreakingHierarchyGalaxy2006, bowerDarkNemesisGalaxy2017, wangEnvironmentalDependenceMassMetallicity2023, jiangDissectingMassQuenching2025, liPhysicalProcessesCoevolution2025, chenTwophaseModelGalaxy2026, chenCoolCircumgalacticEcho2026, kongCoolerPhasesCircumgalactic2026}; and cold-mode accretion, which delivers gas to the ISM without it being shock heated \citep{keresHowGalaxiesGet2005, dekelColdStreamsEarly2009}.
We stress that $\lambda$ is therefore an effective quantity, and that the model does not separate the contributions of these processes (see \citetalias{wanginflow1} for a detailed discussion).

The cooling efficiency is modelled as a function of halo mass and redshift,
\begin{equation}
  \lambda = \frac{\lambda_0 (1 + z)^{\lambda_z}} {(M_\mh / M_\lambda)^{\lambda_{m_1}} + (M_\mh / M_\lambda)^{\lambda_{m_2}}} .
  \label{eq:lambda}
\end{equation}
The double power law in halo mass accounts for the suppression of the gas supply to the ISM at both ends: by preventive feedback at the low-mass end \citep{daveAnalyticModelEvolution2012, mitraEquilibriumModelConstraints2015, luImportancePreventiveFeedback2017}, and by AGN feedback together with inefficient cooling at the high-mass end \citep{bowerBreakingHierarchyGalaxy2006, crotonManyLivesActive2006}, with $M_\lambda$ setting the halo mass at which the gas supply is most efficient.

The halo mass accretion history is modelled following \cite{wechslerConcentrationsDarkHalos2002} as
\begin{equation}
  M_\mh(z) = M_{\rm h,0}\,\exp\!\left( -\frac{8z}{c_\mh}\right) ,
  \label{eq:mah}
\end{equation}
where $M_{\rm h,0}$ is the halo mass at $z = 0$ and $c_\mh$ is the \textit{halo formation time parameter}.
Haloes with high-$c_\mh$ assemble a given fraction of their final mass earlier than those with low-$c_\mh$.
In this parametrisation the shape of the accretion history is set by $c_\mh$ alone, with $M_{\rm h, 0}$ setting the normalisation.

The star formation efficiency is modelled as a function of stellar mass and redshift,
\begin{equation}
  \epsilon = \epsilon_0 \left(\frac{M_\star}{M_\epsilon}\right)^{\epsilon_m} (1 + z)^{\epsilon_z} ,
  \label{eq:epsilon}
\end{equation}
and the mass-loading factor as a function of stellar mass alone,
\begin{equation}
  \eta = \eta_0 \left(\frac{M_\star}{M_\eta}\right)^{\eta_m} .
  \label{eq:eta}
\end{equation}
We calibrated the model parameters in \citetalias{wanginflow1} against three observables: (i) the mass--metallicity relation, (ii) the star-forming main sequence of \citet{jainUniformAnalysisGasphase2026}, and (iii) the stellar-to-halo mass relation of \citet{behrooziUNIVERSEMACHINECorrelationGalaxy2019}. The resulting best-fit values are listed in Table~\ref{tab:parameters}. We next investigate how the scaling relations respond when the parameters are varied about these values.

\subsection{Response of three scaling relations to the input parameters}
\label{sub:response} 

\begin{figure*}
  \begin{center}
    \includegraphics[width=0.9\linewidth]{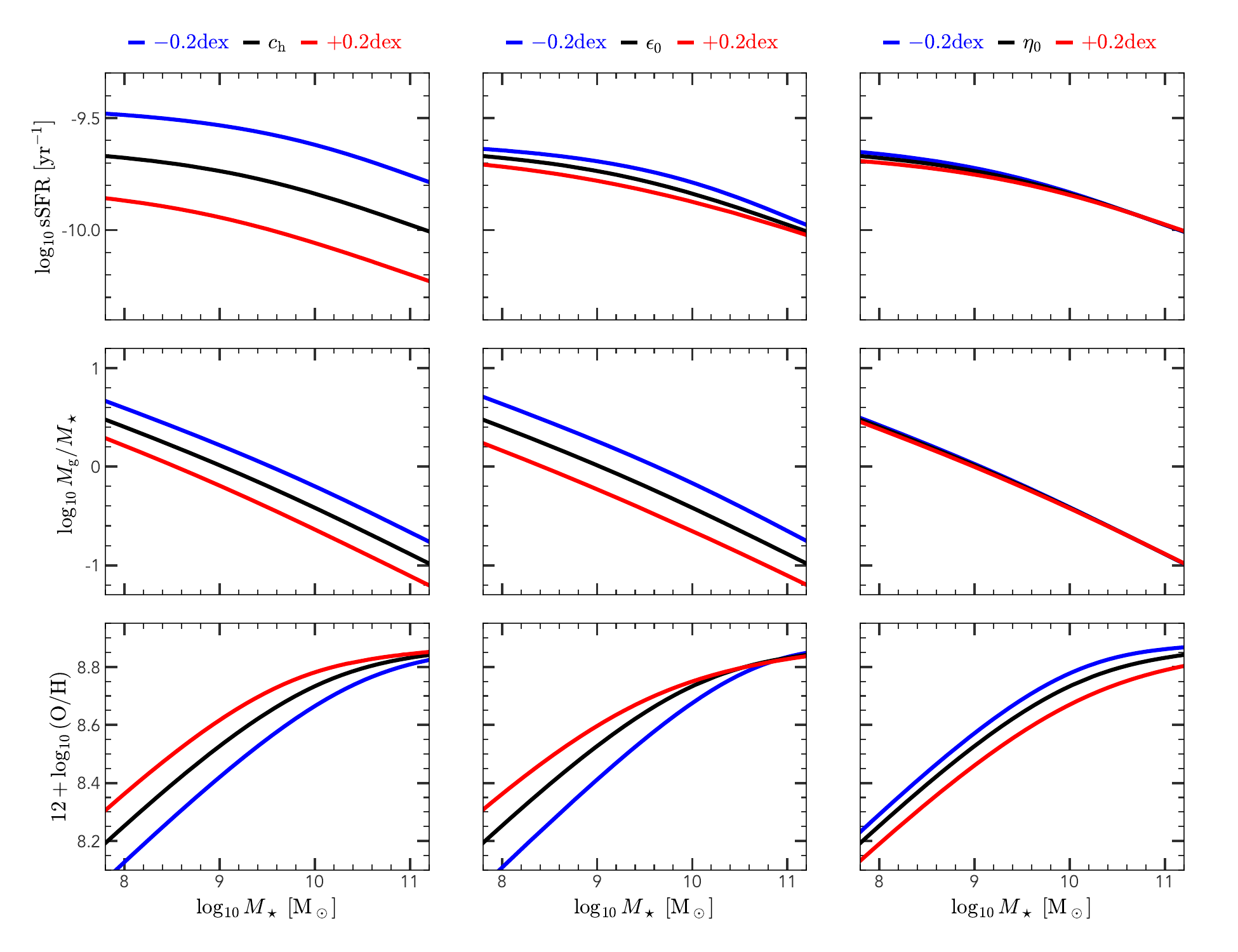}
  \end{center}
  \caption{
    Response of the SFMS ($M_\star$--$\mathrm{sSFR}$; upper row), the gas-to-stellar mass relation ($M_\star$--$M_\mg/M_\star$; centre row), and the MZR ($M_\star$--$Z_\mg$; lower row) to a constant offset in each model parameter at $z = 0$.
    Each column shifts a single parameter by $\pm 0.2$~dex and holds the other two at their fiducial values: the halo formation time $c_\mathrm{h}$ (left), the normalisation of the star formation efficiency $\epsilon_0$ (middle), and the normalisation of the mass-loading factor $\eta_0$ (right).
    The black curve in each panel is the fiducial model result.
    The SFMS responds mainly to halo formation time: an increase in $c_\mathrm{h}$ lowers the sSFR by a nearly constant amount over the whole stellar mass range.
    The response to $\epsilon_0$ is weaker, and the response to $\eta_0$ is negligible.
    The gas fraction responds to $c_\mathrm{h}$ and $\epsilon_0$ by comparable amounts, an increase in either reducing $M_\mg/M_\star$, and is likewise insensitive to $\eta_0$.
    The gas metallicity responds to all three parameters, with a mass dependence that differs between them: $c_\mathrm{h}$ and $\epsilon_0$ shift the low-mass end and leave the massive end nearly unchanged, whereas $\eta_0$ sets the offset at $M_\star \ga 10^{10}\,\Msun$, where an increase in $\eta_0$ lowers $Z_\mg$.
  }
  \label{fig:cos_gas_const_offset}
\end{figure*}

Here we vary the parameters of the cosmological gas flow model about their calibrated values and examine how the three scaling relations respond.
Fig.~\ref{fig:cos_gas_const_offset} shows the three scaling relations at $z = 0$: the SFMS ($M_\star$--$\mathrm{sSFR}$), the gas-to-stellar mass relation ($M_\star$--$M_\mg/M_\star$), and the MZR ($M_\star$--$Z_\mg$).

In each column a single parameter is varied while the other two are held at their fiducial values.
From left to right, the columns vary the halo formation time $c_\mathrm{h}$, the normalisation of the star formation efficiency $\epsilon_0$, and the normalisation of the mass-loading factor $\eta_0$.
The two offset curves in each column show the model with that parameter increased and decreased by $0.2$~dex.

The top row shows the results for the SFMS ($M_\star$--$\mathrm{sSFR}$).
The fiducial relation depends weakly on stellar mass: the sSFR declines by about $0.3$ dex from $10^{8}$ to $10^{11}\,\Msun$, corresponding to a slope of $\approx -0.1$.
Two mechanisms combine to produce this mass dependence.
First, low-mass galaxies sit closer to the inflow-driven regime, in which the ideal model gives a sSFR a factor of two higher than in equilibrium.
Second, the cooling efficiency is suppressed in massive haloes.
This lengthens the time over which their galaxies build stellar mass and hence lowers the sSFR.

Of the three parameters, the SFMS responds most strongly to the halo formation time $c_\mh$, and only weakly to the star formation efficiency and the mass-loading factor.
First, raising $c_\mh$ by $0.2$~dex, which makes haloes form earlier, lowers the amplitude of the SFMS by $\approx 0.2$~dex, and lowering it raises the SFMS by a similar amount.
At fixed stellar mass, a galaxy in an early-formed halo has had longer to form that stellar mass and therefore requires a lower average SFR.
Consequently, its sSFR at the epoch of observation is lower.
Second, the SFMS also responds to the star formation efficiency amplitude ($\epsilon_0$), by a smaller amount and mainly at the low-mass end.
This follows from the equilibrium timescale ($\tau_{\rm eq}\propto 1 / \epsilon$): raising the star formation efficiency shortens $\tau_{\rm eq}$ and brings more galaxies closer to equilibrium, which lowers the sSFR.
Finally, altering the mass-loading factor amplitude ($\eta_0$) has negligible impact on the SFMS.
The mass-loading factor regulates the equilibrium timescale through $\tau_{\rm eq} \propto 1/(1-R+\eta)$, and with our fiducial $\eta \approx 0.88$ at $M_\star \approx 10^8\,\Msun$, a 0.2\,dex change in $\eta_0$ shifts $\tau_{\rm eq}$ by roughly 30 per cent.
The SFMS barely responds because $\eta$ enters the sSFR only through $\mathcal{K}_2'(t/\tau_{\rm eq})$ in equation~(\ref{eq:ssfr}), which is bounded between 1/2 and 1 and varies only slowly with $t/\tau_{\rm eq}$ for galaxies near the inflow-driven regime.

The middle row shows the gas-to-stellar mass relation ($M_\star$--$M_\mg/M_\star$), which follows from the identity $M_\mg/M_\star = \mathrm{sSFR}/\epsilon$.
Its slope is therefore the difference between the slope of the $M_\star$--$\mathrm{sSFR}$ relation ($\approx -0.1$) and that of the $M_\star$--$\epsilon$ relation ($=0.33$), giving $\approx -0.43$.

The response to halo formation time is inherited unchanged from the SFMS: raising (lowering) $c_\mathrm{h}$ by $0.2$~dex lowers (raises) the gas fraction by $\approx 0.2$~dex.
The difference is that the gas fraction also responds to the star formation efficiency, and by a comparable amount.
Raising (lowering) $\epsilon_0$ by $0.2$~dex lowers (raises) the gas fraction by $\ga 0.2$~dex, with the excess coming from the accompanying reduction in the sSFR.
The mass-loading factor again has a negligible effect.

The bottom row shows the MZR ($M_\star$--$Z_\mg$).
The $\kk$ framework (equations~\ref{eq:master}--\ref{eq:t2tau}) sets the gas metallicity from gas fraction alone in the inflow-driven regime ($t\ll \tau_{\rm eq}$), from the mass-loading factor alone in the equilibrium regime ($t \gg \tau_{\rm eq}$), and from both in between.
The gas fraction in turn follows from the SFMS, and hence from the halo formation time, modulated by the star formation efficiency.
We therefore expect low-mass galaxies, whose higher gas fractions place them closer to the inflow-driven regime, to respond mainly to offsets in the halo formation time and in the star formation efficiency, and massive galaxies to respond mainly to the mass-loading factor.
This is what the bottom row shows.
Raising (lowering) either $c_\mathrm{h}$ or $\epsilon_0$ by $0.2$~dex raises (lowers) the gas metallicity by $\approx 0.1$~dex at $M_\star \approx 10^{8}\,\Msun$, and both enter by reducing (increasing) the gas fraction, which raises (lowers) $Z_\mg$ through the inverse dependence in equation~\eqref{eq:master}.
The response declines with increasing stellar mass and becomes negligible at $M_\star \approx 10^{11}\,\Msun$.
The mass-loading factor behaves in the opposite sense, its response strengthening towards the massive end: a $0.2$~dex increase (decrease) in $\eta_0$ lowers (raises) the gas metallicity by $\approx 0.05$~dex.

Finally, we note that the SFMS and the gas-to-stellar mass relation are insensitive to where a galaxy sits in the transition from the inflow-driven regime to equilibrium.
The transition enters both relations only through $\mathcal K_2^\prime$ in equation~\eqref{eq:ssfr}, and hence can move them by at most a factor of two.
The MZR is more sensitive to the transition between the inflow-driven regime and equilibrium.
In our fiducial calibration, galaxies lie closer to the inflow-driven limit, so the star formation efficiency has a larger effect on the relation than the mass-loading factor.
Were galaxies well into equilibrium, the gas metallicity would reduce to $Z_\mg = y/(1 - R + \eta)$, the gas fraction would drop out entirely, and the mass-loading factor would set the relation on its own.

\subsection{The scatter in the scaling relations}
\label{sub:the_scatter_in_the_scaling_relations} 

\begin{figure*}
  \begin{center}
    \includegraphics[width=0.9\linewidth]{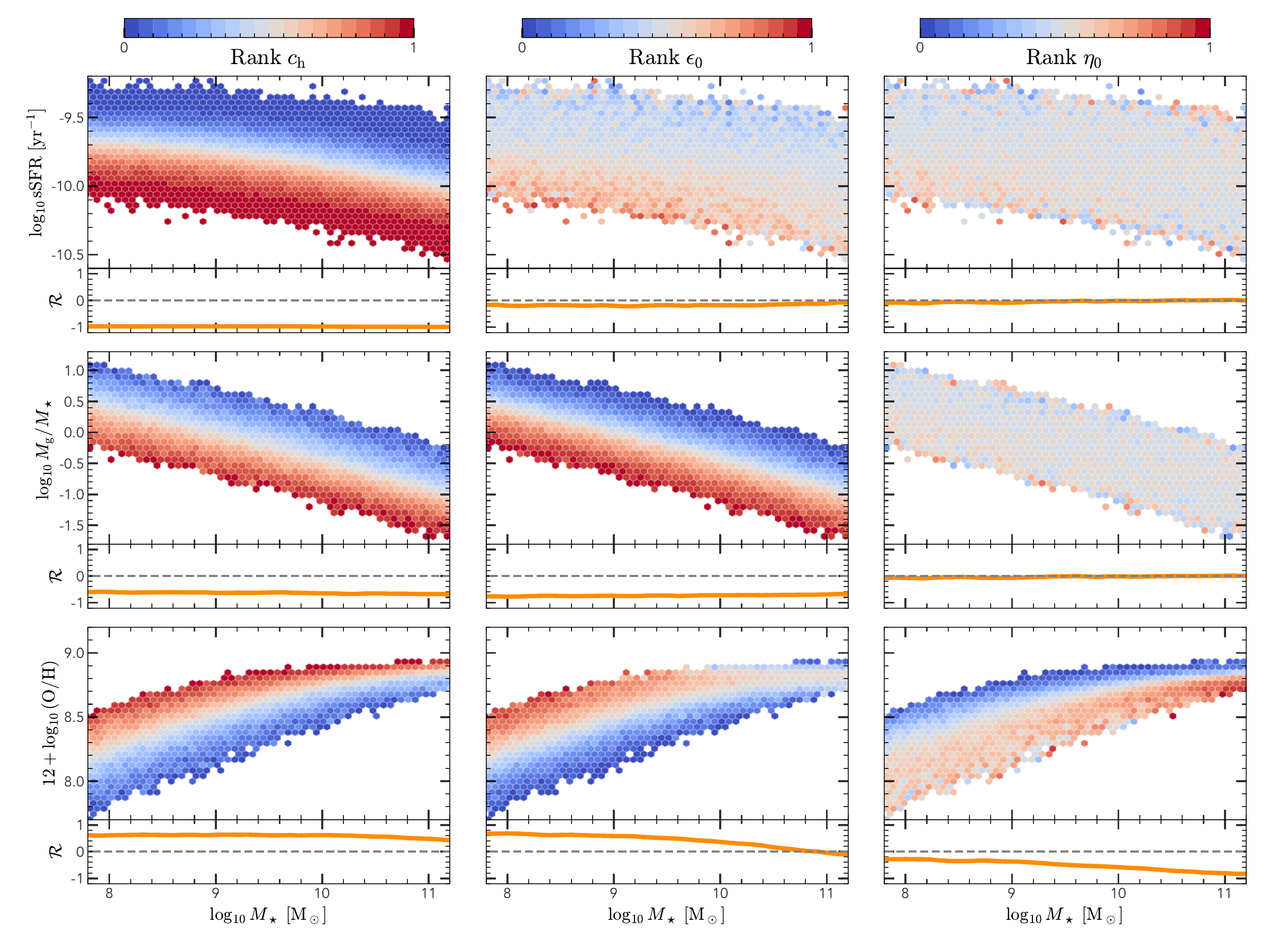}
  \end{center}
  \caption{
    Scatter in the SFMS, the gas-to-stellar mass relation, and the MZR at $z=0$, coloured by the rank of the halo formation time $c_\mathrm{h}$ (left), the star formation efficiency normalisation $\epsilon_0$ (middle), and the mass-loading factor normalisation $\eta_0$ (right).
    Each galaxy is assigned an independent offset drawn from a Gaussian of dispersion $0.15$~dex in each of the three quantities, held fixed along its entire evolutionary track.
    Ranks are computed within bins of stellar mass and normalised to the range $[0,1]$, so that the colour gives the position of a galaxy within the parameter distribution at fixed $M_\star$.
    The narrow sub-panels give the Spearman rank correlation $\mathcal{R}$ between the quantity encoded in colour and the quantity on the $y$-axis at fixed $M_\star$; the dashed line marks zero correlation.
    The scatter of the $M_\star$--$\mathrm{sSFR}$ relation is set almost entirely by halo formation time in this model: an earlier-forming halo has longer to convert its gas into stars and so reaches $z=0$ with a lower sSFR, while neither $\epsilon_0$ nor $\eta_0$ leaves an appreciable imprint.
    The scatter of the $M_\star$--$M_\mg/M_\star$ relation is modulated by halo formation time and star formation efficiency in equal measure, both acting through the rate at which the reservoir is depleted, while mass loading again contributes nothing.
    The scatter of the $M_\star$--$Z_\mg$ relation responds to all three parameters, with their relative weight shifting with mass.
    Halo formation time and star formation efficiency, which together set the gas fraction, both correlate positively: the former at all masses, the latter only below $M_\star \sim 10^{10.5}\,\Msun$.
    Mass loading imposes a negative correlation that strengthens towards high mass.
  }
  \label{fig:correlation_cos_gas_flow}
\end{figure*}

\begin{figure}
  \begin{center}
    \includegraphics[width=0.95\linewidth]{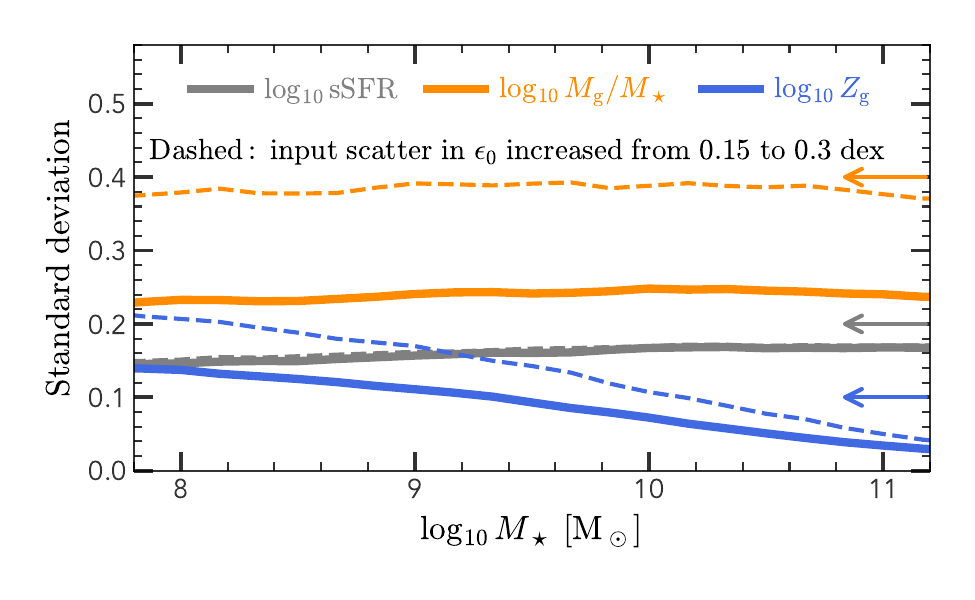}
  \end{center}
  \caption{
    Scatter of the three scaling relations at $z = 0$ as a function of stellar mass, measured as the standard deviation of $\log_{10}{\rm sSFR}$, $\log_{10} M_\mg/M_\star$, and $\log_{10} Z_\mg$ in bins of stellar mass for the population of Fig.~\ref{fig:correlation_cos_gas_flow}, with an input scatter of 0.15~dex in each parameter (solid lines).
    The scatter of the gas fraction exceeds that of the sSFR at all stellar masses, since the identity $M_\mg/M_\star = {\rm sSFR}/\epsilon$ adds the scatter of the star formation efficiency to that inherited from the SFMS.
    The scatter of the gas metallicity is smaller than both and declines with stellar mass, because the mapping from the gas fraction and the mass-loading factor onto the metallicity compresses the inherited scatter, increasingly so as galaxies approach equilibrium.
    Dashed lines show the same population with the input scatter in $\epsilon_0$ increased from 0.15 to 0.3~dex, which raises the scatter of the gas fraction to the observed value while leaving that of the sSFR unchanged.
    The arrows on the right mark the typical observed scatter of each relation, $\approx 0.2$~dex for the SFMS \citep{speagleHighlyConsistentFramework2014}, $\approx 0.4$~dex for the gas fraction \citep{catinellaXGASSTotalCold2018, saintongeColdInterstellarMedium2022a}, and $\la 0.1$~dex for the MZR \citep{tremontiOriginMassMetallicityRelation2004, curtiMassmetallicityFundamentalMetallicity2020}.
    The ranking of the observed values matches that of the model.
  }
  \label{fig:sca_budget}
\end{figure}

\begin{figure*}
  \begin{center}
    \includegraphics[width=0.8\linewidth]{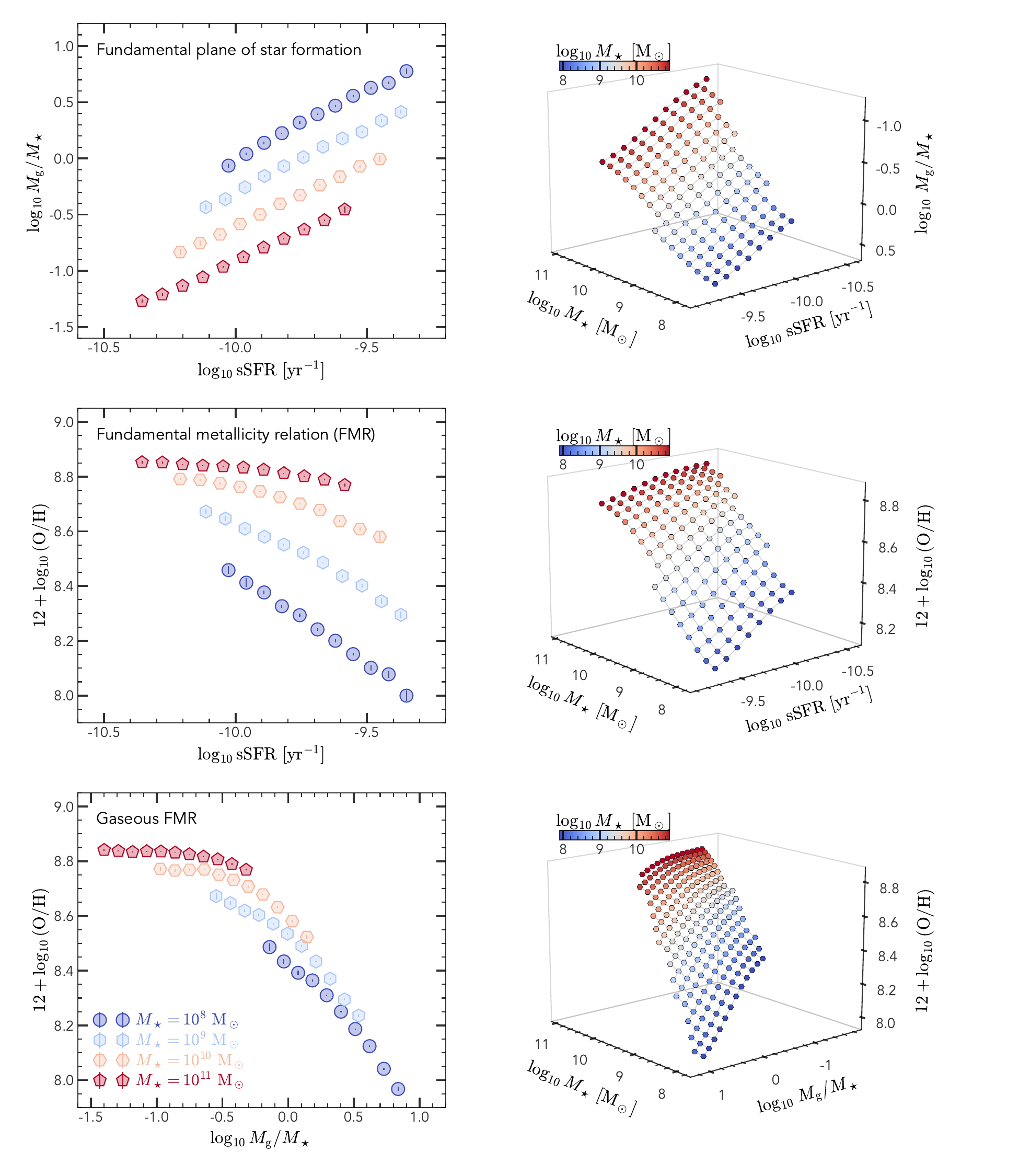}
  \end{center}
  \caption{
    The higher-dimensional scaling relations of the model population of Fig.~\ref{fig:correlation_cos_gas_flow} at $z = 0$.
    Left: the median of the vertical-axis quantity in bins of the horizontal-axis quantity, for galaxies at four fixed stellar masses.
    Right: the corresponding three-dimensional surfaces, showing the median of the vertical-axis quantity in bins of stellar mass and of the second horizontal-axis quantity.
    Top: the fundamental plane of star formation, relating stellar mass, sSFR, and gas fraction.
    Middle: the FMR, relating stellar mass, sSFR, and gas metallicity.
    Bottom: the gaseous FMR, relating stellar mass, gas fraction, and gas metallicity.
    The fundamental plane of star formation is a flat surface, appearing at fixed stellar mass as parallel relations offset by stellar mass.
    The FMR is a curved surface, steep at low stellar mass, where galaxies lie closer to the inflow-driven regime, and flat at high stellar mass, where galaxies approach equilibrium.
    The gaseous FMR nearly collapses onto a single relation for all stellar masses, since the gas fraction already absorbs the stellar mass dependence of the star formation efficiency.
  }
  \label{fig:obs_correlations}
\end{figure*}

We have established how the halo formation time, the star formation efficiency, and the mass-loading factor displace the three scaling relations.
We now ask how variance in each of these parameters contributes to the scatter of these scaling relations.
For this purpose, we generate $20\,000$ galaxy evolution tracks, assigning each an independent random offset simultaneously in the halo formation time $c_\mh$, the normalisation of the star formation efficiency $\epsilon_0$, and the normalisation of the mass-loading factor $\eta_0$.
The offsets are drawn from Gaussian distributions with a dispersion of 0.15~dex.

Fig.~\ref{fig:correlation_cos_gas_flow} shows the three scaling relations, which are the SFMS, the gas-to-stellar mass relation, and the MZR from top to bottom rows.
The columns correspond to the three input parameters, from left to right the halo formation time, the star formation efficiency normalisation, and the mass-loading factor normalisation.
Each is used for the colour coding and for the Spearman rank correlation.
In each panel the colour encodes the rank (from 0 to 1) of the parameter computed within bins of width $0.1$~dex in stellar mass.
The smaller panel beneath each large panel shows Spearman's rank correlation between the quantity on the $y$-axis and the colour-coded parameter, evaluated as a function of stellar mass.

The top row shows that the scatter of the SFMS has a rank correlation with $c_\mh$ of nearly $-1$, so that in this model the scatter is set entirely by the halo formation time.
A correlation of this strength follows from our smooth modelling of the halo accretion history, and hence of the gas inflow history that follows from it.
A real galaxy population also carries short-timescale fluctuations in the SFR, which are absent here by construction and which would reduce the correlation.
The available evidence suggests that this omission does not dominate the scatter.
\citet{mattheeOriginScatterStar2019} found in the EAGLE simulation that $70$ per cent of the SFR variation arises on timescales longer than $2$~Gyr, and that the correlation between the sSFR and the halo half-mass time is about $-0.45$ for galaxies with $10^{9} < M_\star/\Msun < 10^{10}$.
The formation time therefore remains a major driver of the SFMS scatter in hydrodynamical simulations \citep[see also][]{blankNIHAOXXVINature2021}.

The step from the SFMS to the gas-to-stellar mass relation follows from the identity $M_\mg/M_\star = \mathrm{sSFR}/\epsilon$.
The scatter of the gas-to-stellar mass relation is therefore inherited from the scatter of the SFMS, whose scatter primarily comes from variance in the halo formation time, with an additional contribution from variance in the star formation efficiency entering at this step.
The middle row of Fig.~\ref{fig:correlation_cos_gas_flow} confirms this expectation, with both $c_\mh$ and $\epsilon_0$ showing a rank correlation of $\approx -0.7$ with the gas fraction at fixed stellar mass.

The bottom row of Fig.~\ref{fig:correlation_cos_gas_flow} shows the result for the MZR.
First, higher $c_\mh$ and higher $\epsilon_0$ both reduce $M_\mg/M_\star$ and therefore raise $Z_\mg$, so both correlate positively with the gas metallicity ($\mathcal R\approx 0.7$).
Both correlations weaken with increasing stellar mass.
The mass-loading factor correlates negatively with the gas metallicity, with the correlation coefficient strengthening from $\mathcal R\approx -0.2$ at $M_\star\approx 10^8\,\Msun$ to $\mathcal R\approx -0.9$ at $M_\star\approx 10^{11}\,\Msun$.
The $\kk$ framework accounts for both trends: in the inflow-driven regime the gas metallicity is set by the gas fraction, and in the equilibrium regime it is set by the mass-loading factor.
Low-mass galaxies sit close to the inflow-driven regime and high-mass galaxies approach equilibrium, so the sensitivity transfers from $c_\mh$ and $\epsilon_0$ to $\eta_0$ as stellar mass increases.

Having identified which parameter drives the scatter of each relation, we now quantify the amount of that scatter in Fig.~\ref{fig:sca_budget}.
The SFMS has a scatter of $\approx 0.15$--$0.17$~dex, close to the 0.15~dex input scatter in $c_\mh$, which maps onto the sSFR almost one-to-one (see also Fig.~\ref{fig:cos_gas_const_offset}).
The gas fraction has a larger scatter of $\approx 0.23$--$0.25$~dex at all stellar masses.
This follows from the identity $\log_{10}(M_\mg/M_\star) = \log_{10}{\rm sSFR} - \log_{10}\epsilon$, under which the variances of the sSFR and of the star formation efficiency add.
Independent variations would give $\approx 0.22$~dex, and the slight excess arises because a higher $\epsilon$ also lowers the sSFR (\S\,\ref{sub:response}), so the $\rm sSFR$ and $\epsilon$ are anti-correlated.

The gas metallicity behaves differently.
It receives scatter from both the gas fraction and the mass-loading factor, yet its scatter is smaller than that of the gas fraction at all stellar masses, declining from $\approx 0.14$~dex at $M_\star \approx 10^8\,\Msun$ to $\approx 0.03$~dex at $M_\star \approx 10^{11}\,\Msun$.
The mapping from the gas fraction onto the metallicity is non-linear the slope ${\rm d}\log_{10} Z_\mg/{\rm d}\log_{10}(M_\mg/M_\star)$ changes from $-1$ in the inflow-driven regime to zero in equilibrium (equations~\ref{eq:master}--\ref{eq:t2tau}), so the inherited scatter is compressed, increasingly so towards high stellar mass.
The scatter in the mass-loading factor is likewise compressed, since in equilibrium $Z_\mg = y/(1-R+\eta)$ responds to a change in $\eta$ only in proportion to $\eta/(1-R+\eta) \approx 0.16$ at $M_\star \approx 10^{11}\,\Msun$,\footnote{This follows from ${\rm d}\log_{10} Z_\mg/{\rm d}\log_{10}\eta = -\eta/(1-R+\eta)$, evaluated with the parameters of Table~\ref{tab:parameters}.} which turns the 0.15~dex input scatter into $\approx 0.025$~dex, consistent with the $\approx 0.03$~dex in Fig.~\ref{fig:sca_budget}.
The hierarchy therefore implies that scatter is inherited along the chain, but not that it grows at each step.

The amplitudes in Fig.~\ref{fig:sca_budget} scale with the input scatter, which we chose to be 0.15~dex for illustration, and the model omits short-timescale fluctuations and measurement errors.
The robust predictions are the ordering and the mass dependence: the gas fraction should scatter more than the SFMS, and the MZR should be the tightest of the three, with a scatter that increases towards low stellar mass.
Observations follow this pattern (arrows in Fig.~\ref{fig:sca_budget}).
The SFMS has an intrinsic scatter of $\approx 0.2$~dex \citep{speagleHighlyConsistentFramework2014}, and the gas fraction of star-forming galaxies scatters by $\approx 0.4$~dex at fixed stellar mass \citep{catinellaXGASSTotalCold2018, saintongeColdInterstellarMedium2022a}.
The MZR has a scatter of $\la 0.1$~dex, which increases towards low stellar mass \citep{tremontiOriginMassMetallicityRelation2004, mannucciFundamentalRelationMass2010, maiolinoReMetallicaCosmic2019, curtiMassmetallicityFundamentalMetallicity2020}.

Conversely, the observed amplitudes can be used to constrain the scatter of the input parameters.
With the illustrative input of 0.15~dex, the model reproduces the observed scatter of the SFMS and the MZR, but underpredicts that of the gas fraction ($\approx 0.24$ against $\approx 0.4$~dex).
The hierarchy locates the source of this difference: the gas fraction is the only one of the three relations at which the star formation efficiency enters directly, so its scatter is the one that constrains the spread in $\epsilon$.
Increasing the input scatter in $\epsilon_0$ to 0.3~dex raises the scatter of the gas fraction to $\approx 0.4$~dex, in agreement with observations (dashed lines in Fig.~\ref{fig:sca_budget}).
The scatter of the sSFR is unchanged, as expected since $\epsilon$ enters the sSFR only through the transition between the inflow-driven and equilibrium regimes.
The scatter of the MZR increases mainly at low stellar mass, where the metallicity follows the gas fraction, and remains $\la 0.1$~dex above $M_\star \approx 10^{10}\,\Msun$.
The observed scatters therefore indicate a spread of $\approx 0.15$~dex in the halo formation time, consistent with the scatter of halo concentration (\S\,\ref{sub:sfms}), and of $\approx 0.3$~dex in the star formation efficiency.

Beyond the amount of scatter, the correlations among the three observables at fixed stellar mass provide a further test of the framework.
We use the same population as in Figs.~\ref{fig:correlation_cos_gas_flow} and \ref{fig:sca_budget}, in which each galaxy is assigned independent random offsets in $c_\mh$, $\epsilon_0$, and $\eta_0$, held fixed along its evolutionary track, so the correlations arise only from how these offsets propagate through the hierarchy.
Fig.~\ref{fig:obs_correlations} shows these correlations for the model population, in bins of stellar mass.
The top panels show the gas fraction against the sSFR, which at fixed stellar mass is equivalent to the fundamental plane of star formation \citep[$M_\star$--$M_\mg$--SFR;][]{lagosFundamentalPlaneStar2016}.
In each stellar mass bin the gas fraction rises with the sSFR with a slope close to unity, as expected from the identity $M_\mg/M_\star = {\rm sSFR}/\epsilon$, and the relations for different stellar masses are parallel and offset by the stellar mass dependence of $\epsilon$, so the galaxies occupy a plane in the three-dimensional space.

The middle panels show the gas metallicity against the sSFR, which at fixed stellar mass is the FMR \citep{mannucciFundamentalRelationMass2010}.
This surface is curved rather than planar: the gas metallicity declines steeply with the sSFR at low stellar mass, and becomes nearly independent of the sSFR at high stellar mass.
The same behaviour is found in observations \citep{mannucciFundamentalRelationMass2010, curtiMassmetallicityFundamentalMetallicity2020}.
As shown in \citetalias{wanginflow1}, this transition follows from the evolutionary state of galaxies.
Low-mass galaxies lie closer to the inflow-driven regime, in which $Z_\mg = [y/(1-R)]\,(M_\mg/M_\star)^{-1} \propto {\rm sSFR}^{-1}$ at fixed $\epsilon$, so the slope approaches $-1$.
Massive galaxies lie closer to equilibrium, in which $Z_\mg = y/(1-R+\eta)$ is set by the mass-loading factor alone, so the gas metallicity becomes independent of the sSFR.
The gas regulator model of \citet{lillyGASREGULATIONGALAXIES2013}, which assumes the gas reservoir to be in quasi-equilibrium, does not reproduce this transition \citepalias[see][for a detailed discussion]{wanginflow1}.

The bottom panels show the gas metallicity against the gas fraction, which at fixed stellar mass is the gaseous FMR \citep{bothwellFundamentalRelationMetallicity2013, jimmyGasPhaseMass2015}.
Its shape mirrors that of the FMR, steep at high gas fraction and flat at low gas fraction, but the relations for different stellar masses nearly collapse onto one.
This is because the gas metallicity is set by the gas fraction and the mass-loading factor alone through the $\kk$ framework.
The small residual offsets at low gas fraction reflect the stellar mass dependence of $\eta$, as massive galaxies have a lower mass-loading factor and so reach a higher equilibrium metallicity.
The tighter dependence of the gas metallicity on the gas fraction than on the sSFR is consistent with observations \citep{bothwellFundamentalRelationMetallicity2013, jimmyGasPhaseMass2015}.

\subsection{A unified picture of the scaling relations}
\label{sub:a_unified_picture} 

\begin{figure*}
  \begin{center}
    \includegraphics[width=\linewidth]{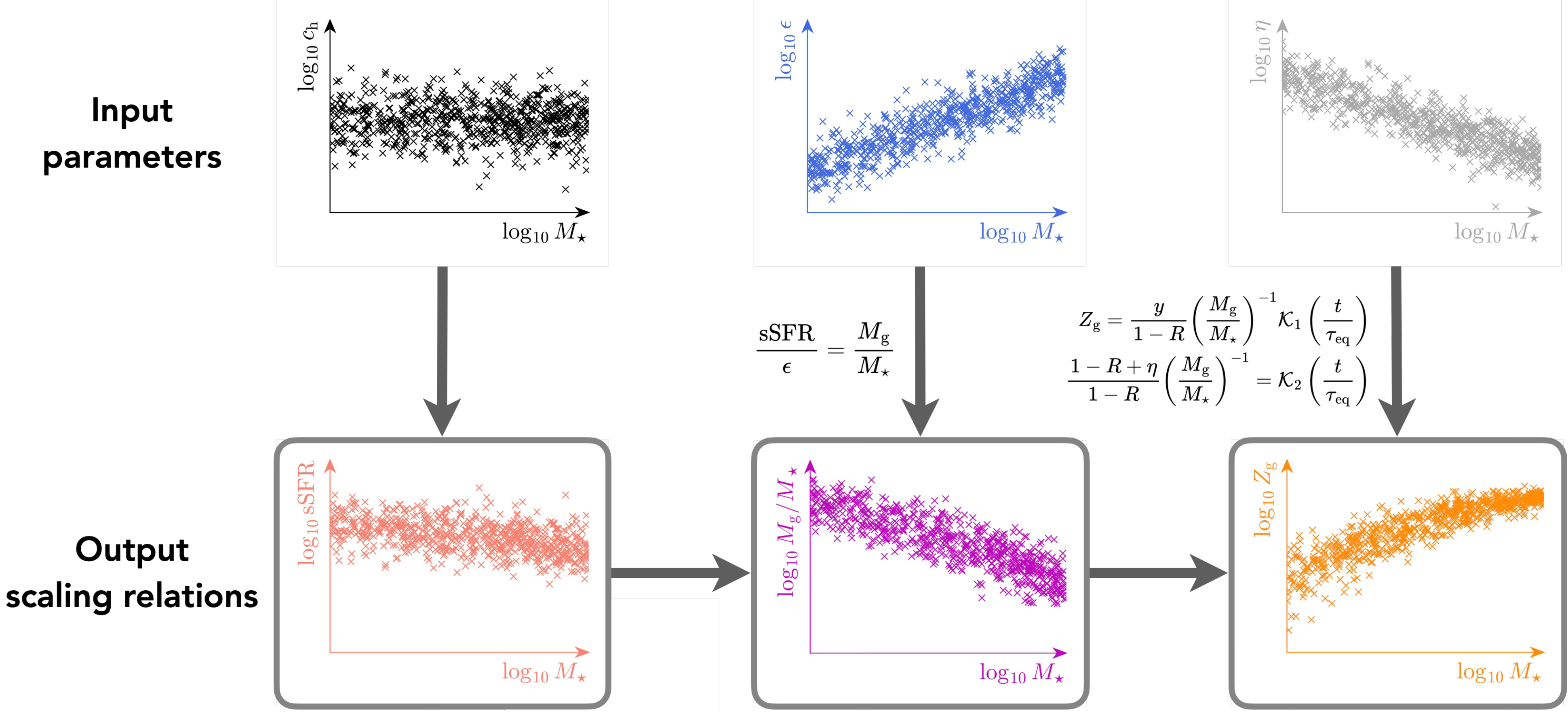}
  \end{center}
  \caption{
    Schematic of how the parameters of the gas flow equations, the halo formation time parameter $c_\mathrm{h}$, the star formation efficiency $\epsilon$, and the mass-loading factor $\eta$, build the three scaling relations and their scatter.
    Upper row: the three input parameters against stellar mass, each scattered about its median trend.
    Lower row: the resulting sSFR, gas fraction, and gas-phase metallicity, obtained by propagating the inputs through the relations written between the panels.
    The arrows trace the order in which the parameters enter.
    Halo formation time primarily sets the sSFR; $\epsilon$ then converts the sSFR into a gas fraction through $\mathrm{sSFR}/\epsilon = M_\mathrm{g}/M_\star$; $\eta$ enters last, mapping the gas fraction onto the metallicity through the analytic $\mathcal{K}_1$--$\mathcal{K}_2$ equations.
    The scatter therefore accumulates along the chain: the main sequence inherits scatter from one parameter, the gas fraction relation from two, and the MZR from all three.
  }
  \label{fig:demo}
\end{figure*}

\begin{figure}
  \begin{center}
    \includegraphics[width=0.95\linewidth]{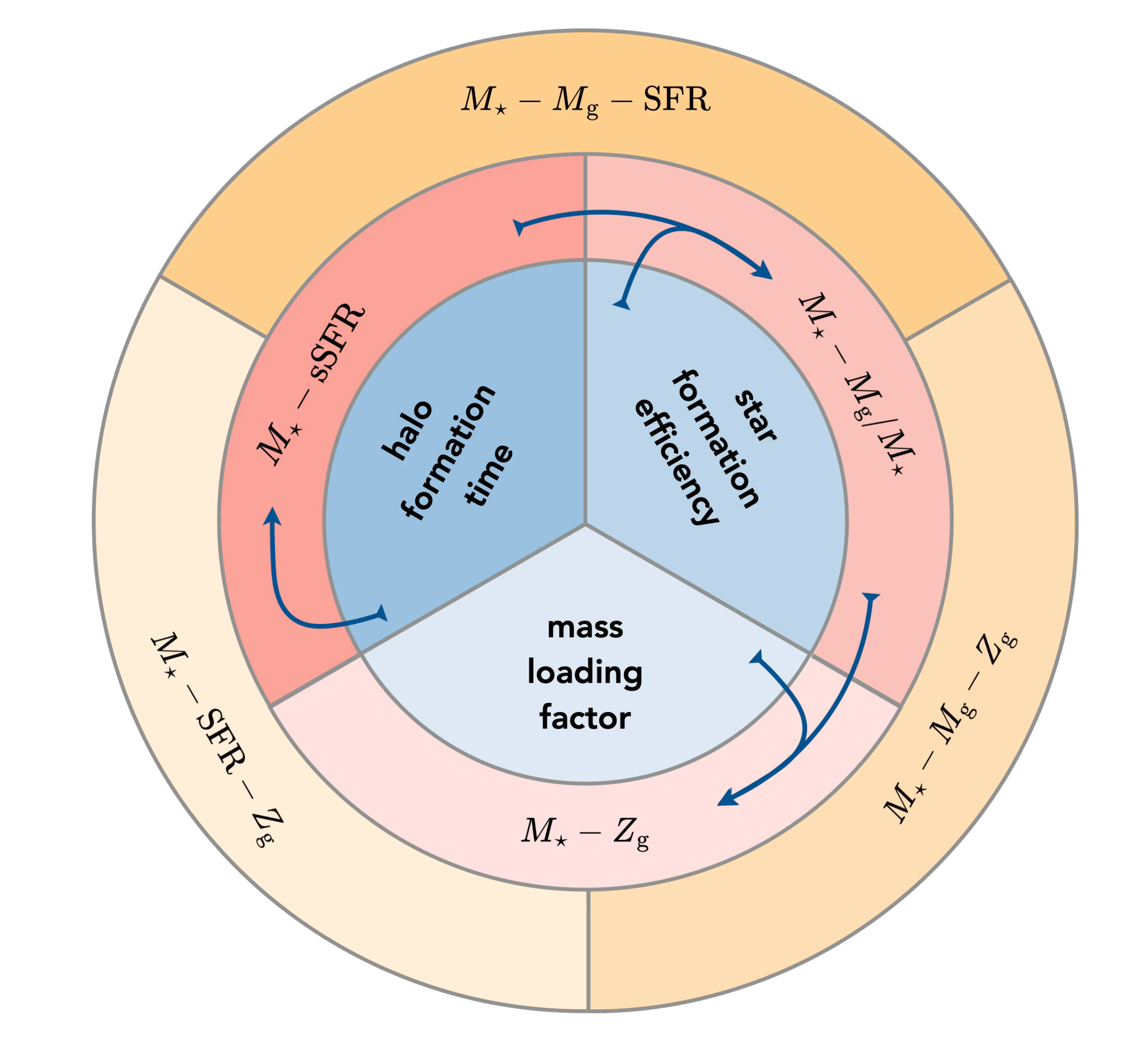}
  \end{center}
  \caption{
    A unified framework for the scaling relations of star-forming galaxies.
    The innermost circle shows the three physical parameters of the gas flow model: the halo formation time, the star formation efficiency, and the mass-loading factor.
    The middle ring shows the three primary scaling relations they set: the SFMS ($M_\star$--sSFR), the gas-to-stellar mass relation ($M_\star$--$M_{\rm g}/M_\star$), and the MZR ($M_\star$--$Z_{\rm g}$).
    The arrows trace the hierarchy among them (see also Fig.~\ref{fig:demo}).
    The outermost ring shows the higher-dimensional scaling relations, each spanning the two primary relations it connects: the fundamental plane of star formation ($M_\star$--$M_{\rm g}$--SFR), the gaseous FMR ($M_\star$--$M_{\rm g}$--$Z_{\rm g}$), and the FMR ($M_\star$--SFR--$Z_{\rm g}$).}
  \label{fig:demo2}
\end{figure}

Using the ideal gas flow model and the more realistic cosmological gas flow model built in \citetalias{wanginflow1}, we have established a hierarchy among the three scaling relations of star-forming galaxies: the SFMS ($M_\star$--$\rm sSFR$), the gas-to-stellar mass relation ($M_\star$--$M_\mg/M_\star$), and the MZR ($M_\star$--$Z_\mg$), which is summarised in Fig.~\ref{fig:demo}.
All three relations are expressed as a function of stellar mass, so throughout we assess scatter at fixed stellar mass.

The SFMS sits at the top of the hierarchy.
Since $\int (1-R){\rm SFR}(t)\dd t = M_\star$, the scatter of the SFMS at fixed stellar mass can only be produced by a process that modulates the timescale over which that stellar mass is formed, provided the star formation history is not dominated by stochastic fluctuations.
Displacing the star formation efficiency or the mass-loading factor does alter the SFR of an individual galaxy, but it alters the stellar mass in the same direction, displacing the galaxy along the SFMS without contributing to the scatter.

The gas-to-stellar mass relation follows from the identity $M_\mg/M_\star = {\rm sSFR}/\epsilon$.
It inherits the scatter of the SFMS and acquires an additional contribution from the variance of star formation efficiency.

The MZR sits one level further down.
The $\kk$ framework of \citetalias{wanginflow1} shows that the gas metallicity is set jointly by the gas fraction and the mass-loading factor: in the inflow-driven limit it depends only on the gas fraction, in the equilibrium limit only on the mass-loading factor, and on both in between.
The MZR therefore inherits the scatter of the two relations above it and acquires a further contribution from the variation in the mass-loading factor, weighted by how close galaxies sit to equilibrium.

Fig.~\ref{fig:demo2} extends the framework to the three higher-dimensional scaling relations.
The innermost circle shows the three parameters that govern the system: the halo formation time, the star formation efficiency, and the mass-loading factor.
Through the hierarchy of Fig.~\ref{fig:demo}, these parameters set the three primary scaling relations in the middle ring.
The outermost ring shows the three higher-dimensional scaling relations (see Fig.~\ref{fig:obs_correlations}), each connecting a pair of primary relations.
The fundamental plane of star formation \citep[$M_\star$--$M_\mg$--SFR;][]{lagosFundamentalPlaneStar2016} adds the gas mass, $M_\mg = {\rm SFR}/\epsilon$, to the SFMS as a third axis.
If the star formation efficiency is a power law in stellar mass \citep{catinellaXGASSTotalCold2018, saintongeColdInterstellarMedium2022a}, $\log_{10} M_\mg$ is a linear combination of $\log_{10} M_\star$ and $\log_{10} {\rm SFR}$, and galaxies lie on a plane in logarithmic space.
The gaseous FMR ($M_\star$--$M_\mg$--$Z_\mg$) follows directly from the $\kk$ framework in equations~\eqref{eq:master}--\eqref{eq:t2tau}, in which the gas metallicity is set by the gas fraction and the mass-loading factor, itself a function of stellar mass.
As shown in \citetalias{wanginflow1}, the FMR \citep[$M_\star$--SFR--$Z_\mg$;][]{mannucciFundamentalRelationMass2010} follows from the gaseous FMR by substituting $M_\mg = {\rm SFR}/\epsilon$, so that the parametrisation of the observed FMR encodes the stellar mass dependence of the star formation efficiency.

\section{The mass and redshift dependence of the scaling relations}
\label{sec:what_shape}

\subsection{The mass-dependence of star formation efficiency and mass-loading factor}
\label{sub:m_sfe_loading} 

\begin{figure}
  \begin{center}
    \includegraphics[width=\linewidth]{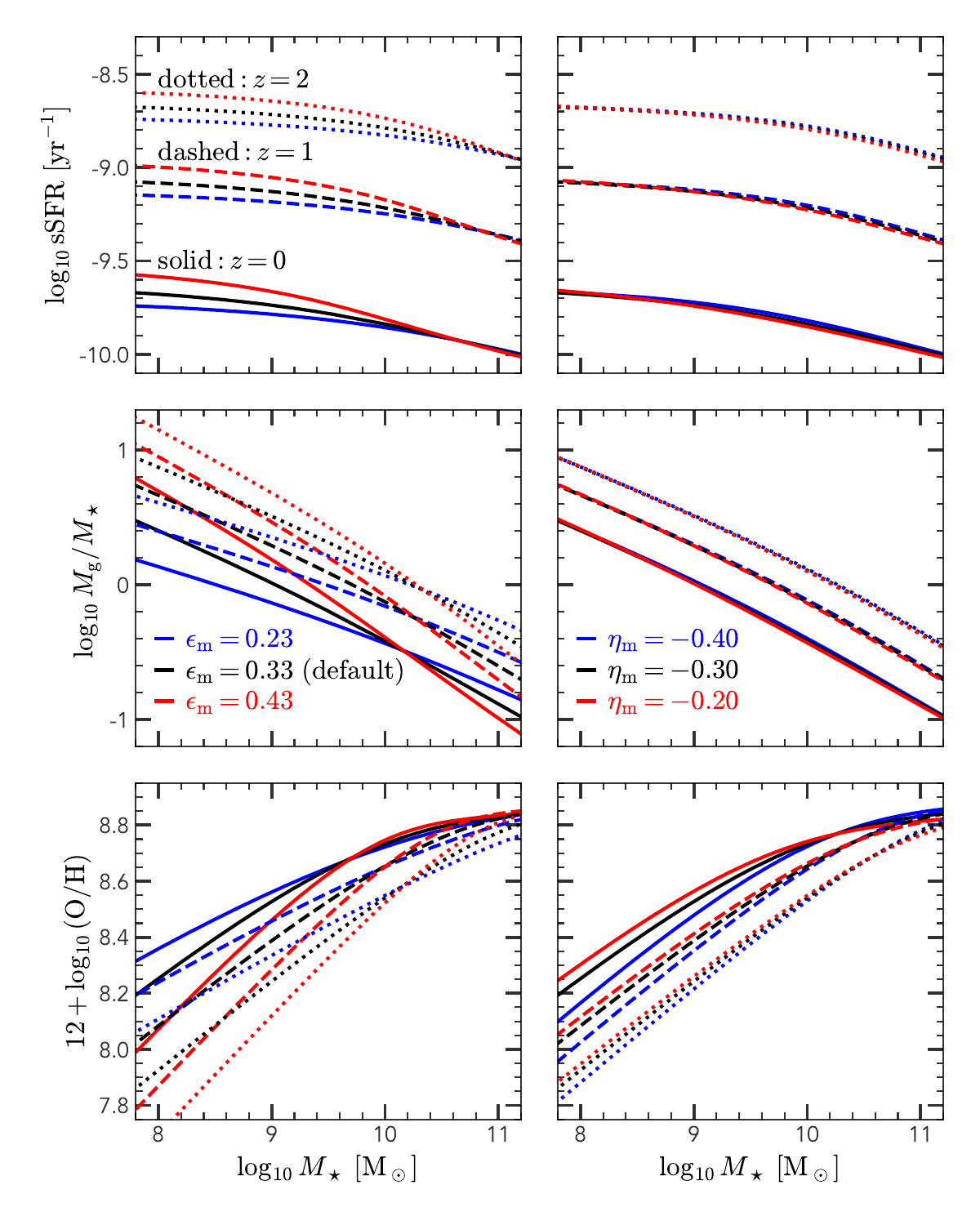}
  \end{center}
  \caption{
    The SFMS (upper panel), the gas-to-stellar mass relation (centre panel), and the MZR (lower panel), and their response to the mass dependence of the star formation efficiency (left) and of the mass-loading factor (right).
    Both prescriptions are anchored at $M_\epsilon = M_\eta = 10^{10}\,\Msun$, so varying either slope pivots the corresponding parameter about that mass.
    Line styles give the redshift, $z=0$ (solid), $z=1$ (dashed), and $z=2$ (dotted); the black curves mark the fiducial slope.
    The main sequence barely moves in either column.
    Varying $\epsilon_\mathrm{m}$ pivots the gas fraction about $M_\star \sim 10^{10}\,\Msun$ and shifts the gas metallicity at the low-mass end, with a larger $\epsilon_\mathrm{m}$ giving a higher gas fraction and a lower gas metallicity there, while the offset curves converge above $M_\star \sim 10^{10}\,\Msun$.
    Varying $\eta_\mathrm{m}$ leaves both the SFMS and the gas fraction unchanged and acts only on the gas metallicity, with a weaker amplitude than $\epsilon_\mathrm{m}$.
  }
  \label{fig:cos_gas_sfe_loading_m}
\end{figure}

We now ask how the stellar mass dependences of the star formation efficiency and the mass-loading factor affect the shape of the three scaling relations.
The mass dependence is governed by $\epsilon_{\rm m}$ and $\eta_{\rm m}$, where $\epsilon \propto (M_\star/M_\epsilon)^{\epsilon_{\rm m}}$ and $\eta \propto (M_\star/M_\eta)^{\eta_{\rm m}}$.
Fig.~\ref{fig:cos_gas_sfe_loading_m} shows the relations at $z = 0$, $1$, and $2$, with the left column varying the slope of the star formation efficiency $\epsilon_{\rm m}$ by $\pm 0.1$ around its fiducial value of $0.33$, and the right column varying the slope of the mass-loading factor $\eta_{\rm m}$ by $\pm 0.1$ around its fiducial value of $-0.30$.
Both prescriptions are anchored at $M_\epsilon = M_\eta = 10^{10}\,\Msun$, so a change in either slope pivots the corresponding parameter about that mass and leaves it unchanged there.

The SFMS responds to $\epsilon_{\rm m}$ mainly at the low-mass end.
Raising $\epsilon_{\rm m}$ lowers $\epsilon$ below $M_\epsilon$, which lengthens $\tau_{\rm eq}$ and moves low-mass galaxies towards the inflow-driven regime, and hence raises their sSFR.
Lowering $\epsilon_{\rm m}$ drives the same galaxies closer to equilibrium and lowers their sSFR.
The response is bounded by a factor of two, as established in \S\,\ref{sec:perspective_ideal}: $\epsilon$ enters the sSFR only through the transition between the inflow-driven and equilibrium regimes, whose sSFRs differ by that factor.

The gas-to-stellar mass relation responds more strongly to $\epsilon_{\rm m}$, since $\epsilon$ enters it explicitly through the identity $M_\mg/M_\star = \mathrm{sSFR}/\epsilon$.
Below $M_\epsilon$, raising $\epsilon_{\rm m}$ both raises the sSFR and lowers $\epsilon$, and the two contributions add to raise the gas fraction.
Above $M_\epsilon$ both are reversed and the gas fraction falls.
Lowering $\epsilon_{\rm m}$ produces the opposite response.
The relation therefore pivots about $M_\epsilon$, steepening as $\epsilon_{\rm m}$ increases.

The change in gas fraction propagates into the gas metallicity through the $\kk$ framework, a raised gas fraction lowering $Z_\mg$ through the inverse dependence in equation~\eqref{eq:master}.
The response is largest for low-mass galaxies and at high redshift, in both cases because the higher gas fractions place these galaxies closer to the inflow-driven regime.

The right column shows the corresponding response to $\eta_{\rm m}$.
The SFMS and the gas-to-stellar mass relation are insensitive to it, since $\eta$ enters both only through $\mathcal{K}_2'$ in equation~\eqref{eq:ssfr}, which varies by less than a factor of two, so the MZR is the only relation to respond.
Raising $\eta_{\rm m}$ from $-0.40$ to $-0.20$ lowers $\eta$ below $M_\eta$ and raises it above, which raises $Z_\mg$ at the low-mass end and lowers it at the massive end.

Finally, the response to both slopes is close to identical at all three redshifts at fixed stellar mass, so that changing either slope shifts the relations without altering their redshift dependence.
The one exception is the MZR, which responds slightly more to $\epsilon_{\rm m}$ at high redshift and slightly more to $\eta_{\rm m}$ at low redshift.
This follows from the evolution of the gas fraction, which places high-redshift galaxies closer to the inflow-driven regime, where the metallicity is set by the gas fraction, and low-redshift galaxies closer to equilibrium, where it is set by the mass-loading factor.

\subsection{The redshift-dependence of star formation efficiency}
\label{sub:z_sfe} 

\begin{figure}
  \begin{center}
    \includegraphics[width=0.8\linewidth]{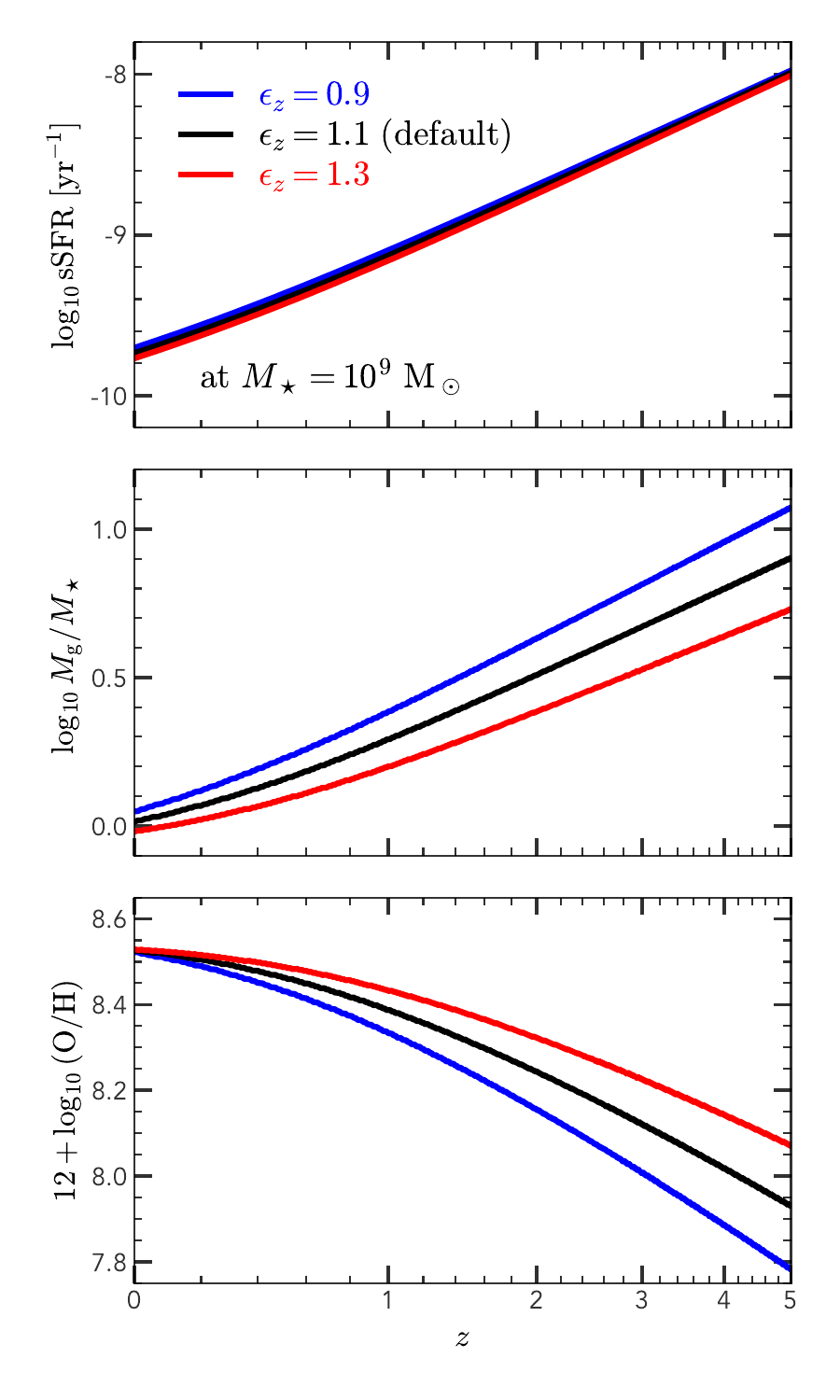}
  \end{center}
  \caption{
    Evolution of the sSFR (top), the gas fraction (middle), and the gas-phase metallicity (bottom) at fixed stellar mass $M_\star = 10^{9}\,\Msun$, and their response to the redshift dependence of the star formation efficiency $\epsilon_z$.
    The black curve marks the fiducial value.
    The sSFR is almost unaffected across the whole range.
    A stronger redshift dependence depletes the gas reservoir faster at early times, lowering the gas fraction by an amount that grows towards high redshift and raising the metallicity there.
    The three metallicity curves converge by $z \sim 0$, so constraining $\epsilon_z$ requires measurements of the MZR at high redshift.
  }
  \label{fig:cos_gas_sfe_z}
\end{figure}

We now turn to the impact of the redshift dependence of the star formation efficiency ($\epsilon_z$, where $\epsilon \propto (1+z)^{\epsilon_z}$) on the three scaling relations.
  Fig.~\ref{fig:cos_gas_sfe_z} shows the evolution of the sSFR, the gas fraction, and the gas metallicity at fixed stellar mass $M_\star = 10^{9}\,\Msun$, for $\epsilon_z$ at its fiducial value of $1.1$ and offset by $\pm 0.2$.
  The comparison is made at fixed stellar mass at each redshift, so the curves do not trace individual galaxies.

  The sSFR at fixed stellar mass increases monotonically with redshift, and this evolution is almost independent of $\epsilon_z$.
  The insensitivity follows from the result of \S\,\ref{sec:perspective_ideal} that $\epsilon$ enters the sSFR only through $\mathcal K_2^\prime$ in equation~\eqref{eq:ssfr}.
  The rise in sSFR with redshift follows from the mass accretion history.
  Writing equation~\eqref{eq:mah} in terms of the scale factor $a = 1/(1+z)$ gives $M_\mh(a) = M_{\rm h,0}\exp[-(8/c_\mh)(1/a - 1)]$, and the growth relative to an arbitrary epoch $a_{\rm o}$ is
  \begin{equation}
    \frac{M_\mh(a)}{M_\mh(a_{\rm o})}
    = \exp\left[-\frac{8}{c_\mh}\left(\frac{1}{a} - \frac{1}{a_{\rm o}}\right)\right]
    = \exp\left[-\frac{8}{c_\mh\,a_{\rm o}}\left(\frac{1}{a'} - 1\right)\right],
    \label{eq:mah_selfsimilar}
  \end{equation}
  where $a_{\rm o} = 1/(1+z_{\rm o})$ is the scale factor at the epoch of observation, $z_{\rm o}$ is the corresponding redshift, and $a' = a/a_{\rm o}$ is the scale factor measured relative to that epoch.
  The history therefore has the same form at every epoch, with the formation time parameter replaced by $c_\mh\,a_{\rm o} = c_\mh/(1+z_{\rm o})$.
  The effective formation parameter $c_\mh/(1+z_{\rm o})$ is therefore smaller at higher redshift, so the same halo is effectively later forming when observed at earlier epochs.
  This reflects the transition from fast to slow accretion \citep{wechslerConcentrationsDarkHalos2002, zhaoGrowthStructureDark2003, ludlowMassProfileAccretion2013}.
  Since later formation raises the sSFR, the sSFR at fixed stellar mass rises with redshift.

  The gas fraction follows from the identity $M_\mg/M_\star = \mathrm{sSFR}/\epsilon$, so its evolution relative to that of the sSFR is governed by $\epsilon_z$ alone.
  A larger $\epsilon_z$ gives a higher star formation efficiency at high redshift and hence a lower gas fraction there, while leaving the $z = 0$ value nearly unchanged.
  The three curves therefore fan out with increasing redshift.

  The change in gas fraction propagates into the gas metallicity through the $\kk$ framework, a lower gas fraction giving a higher $Z_\mg$.
  The response grows with redshift, both because the offset in gas fraction grows and because high-redshift galaxies lie closer to the inflow-driven regime, in which the metallicity is governed by the gas fraction.

  \section{Discussion}
  \label{sec:discussion}

  \subsection{The origin of the star-forming main sequence}
  \label{sub:sfms} 

  The SFMS ($M_\star$--$\mathrm{sSFR}$) is among the most fundamental scaling relations of galaxies \citep{brinchmannPhysicalPropertiesStarforming2004, salimUVStarFormation2007, whitakerStarFormationMass2012, speagleHighlyConsistentFramework2014, popessoMainSequenceStarforming2023}.
  The sSFR is nearly independent of stellar mass below $\sim 10^{10.5}\,\Msun$, above which the relation bends and declines, and the overall amplitude increases steadily with redshift.
  The scatter about the SFMS is typically $0.2$~dex \citep{speagleHighlyConsistentFramework2014}.
  Two features therefore require explanation: the tightness of the relation at fixed redshift, and the steady evolution of its amplitude with cosmic time.

  We focus here on star-forming central galaxies below $10^{10.5}\,\Msun$.
  In this regime the SFR is approximately linearly proportional to stellar mass \citep{popessoMainSequenceStarforming2023}, AGN feedback has not yet taken a central role \citep{bowerDarkNemesisGalaxy2017, wangTestingGalaxyFormation2025}, environmental effects can be neglected \citep{baldryGalaxyBimodalityStellar2006, pengMassEnvironmentDrivers2010}, and galaxies acquire most of their stars through \textit{in situ} star formation rather than through the accretion of satellites \citep{rodriguez-gomezStellarMassAssembly2016, davisonEAGLEsViewEx2020, proctorWeakConnectionStellar2025}.

  One view is that the SFR of a galaxy simply fluctuates randomly about a mean trend, driven by the variations in gas mass and star formation efficiency \citep{forbesOriginFundamentalMetallicity2014, tacchellaConfinementStarformingGalaxies2016, faucher-giguereModelOriginBursty2018, torreySimilarStarFormation2018, caplarAGNEvolutionGalaxy2018, linALMaQUESTSurveyMolecular2019, feldmannLinkStarFormation2020, tacchellaStochasticModellingStarformation2020, wangVariabilityStarFormation2020, wangVariabilityStarFormation2020a, saintongeColdInterstellarMedium2022a}.
  This class of models does not explain the origin of the SFMS, the mean trend that sets the evolution of the sSFR, but instead assumes its existence and models the scatter about it as a stochastic process.
  The cosmic evolution of the SFMS raises a conceptual difficulty for this picture: galaxies across the Universe must either communicate with each other or share a common clock to establish the SFMS and then fluctuate around it \citepalias[a point also made in][]{wanginflow1}.

  Moreover, we can use the observed scatter and redshift evolution of the SFMS to place an upper limit on the timescale over which stochastic variations could dominate the SFR history. In these stochastic fluctuation models, the SFR evolution of a galaxy is decomposed into the systematic evolution of the mean SFMS and stochastic fluctuations around it \citep[e.g.][]{caplarStochasticModellingStarformation2019, tacchellaStochasticModellingStarformation2020}. The change in SFR driven by the systematic evolution of the SFMS grows with the timescale considered, because its normalisation evolves with cosmic time. For example, the SFMS calibration of \citet{speagleHighlyConsistentFramework2014} gives
  \begin{align}
  \log_{10} \left(\frac{\rm SFR}{\Msun\,{\rm yr^{-1}}}\right) & \approx
  \left(0.11 - 0.026 \log_{10} \frac{M_\star}{\Msun}\right) \frac{t}{\rm Gyr}
  \nonumber                                                                                          \\
  & + \text{(terms independent of $t$)},
  \label{eq:speagle_t}
  \end{align}
  so that the normalisation of the SFMS declines at a rate of approximately $0.12$--$0.15$~dex~Gyr$^{-1}$ for galaxies with $M_\star=10^{9}$--$10^{10}\,\Msun$.
  In contrast, the amplitude of the stochastic component is constrained by the intrinsic scatter of the SFMS, which is approximately $0.2$~dex and shows no strong redshift dependence \citep{speagleHighlyConsistentFramework2014}.
  The systematic evolution of the SFMS therefore changes the characteristic SFR by one scatter width in only $0.2/0.15\approx 1.3$  to $0.2/0.12 \approx 1.7$~Gyr over this mass range. Beyond this timescale, the systematic evolution of the SFMS produces a larger change in SFR than stochastic fluctuations confined to its observed scatter. We therefore infer an upper limit of $\approx 1.3$--$1.7$~Gyr on the timescale over which such stochastic variations can dominate the SFR history.

  In contrast to the previous picture, we advocate that the SFMS and its scatter are primarily established by the formation histories of dark matter haloes.
  At fixed stellar mass, galaxies in early-formed haloes have had more time to form their stellar mass, and hence have a lower SFR than galaxies in late-formed haloes, placing them below the SFMS while galaxies in late-formed haloes lie above it.
  This picture is supported by semi-analytic models \citep{boucheImpactColdGas2010, duttonOriginGalaxyStarformationrate2010}, hydrodynamical simulations \citep{mattheeOriginScatterStar2019}, and the observed dependence of galaxy clustering on the offset from the SFMS \citep{bertiMainsequenceScatterReal2021}.

  Physically, a galaxy's star formation history should align closely with the accretion history of its host dark matter halo.
  This alignment arises because star formation is sustained by gas accreted along with halo growth.
  If the two histories were unrelated, the tight stellar-to-halo mass relation observed at $z = 0$ would break down at higher redshift.
  There is no physical basis for such a relation to emerge only at the present epoch.

  Our scenario also accounts for the mass dependence and the redshift evolution of the SFMS.
  The concentration parameter, which strongly correlates with the halo formation time \citep{navarroUniversalDensityProfile1997, wechslerConcentrationsDarkHalos2002}, depends only weakly on halo mass \citep{netoStatisticsCDMHalo2007, zhaoACCURATEUNIVERSALMODELS2009, ludlowMassProfileAccretion2013, wangEfficientRobustMethod2024}, which explains why the sSFR depends only weakly on stellar mass below $M_\star \approx 10^{10.5}\,\Msun$.
  The amplitude of the SFMS declines with cosmic time because halo growth passes from the fast accretion phase to the slow accretion phase \citep{navarroUniversalDensityProfile1997, wechslerConcentrationsDarkHalos2002, zhaoGrowthStructureDark2003}, and because the galaxy evolution time is bounded above by the age of the Universe, which is shorter at higher redshift.
  Moreover, the scatter of the halo concentration is about $0.15$~dex at $z = 0$ for relaxed haloes \citep{wangEfficientRobustMethod2024}, which corresponds to a $\approx 0.15$~dex contribution to the scatter in the SFMS (Fig.~\ref{fig:cos_gas_const_offset}).
  The remainder leaves room for a short-timescale stochastic component, which may well be present although it is not modelled in this work, but which cannot dominate the long-timescale variation, as argued above.

  This scenario further predicts that the evolutionary tracks of individual galaxies are not ergodic about the SFMS, but instead hold a coherent relative position: a galaxy above the SFMS at $z = 0$ is more likely to have been above it in the past, and likewise for galaxies below it.
  This is supported by hydrodynamical simulations: \citet{mattheeOriginScatterStar2019} found that galaxies above the SFMS have on average remained above it for half the age of the Universe.
  In observations, \citet{wanDecodingVariabilityStar2025} found that the coherence of a galaxy's relative position on the SFMS persists for at least $\approx 1$~Gyr.

  \subsection{The origin of the gas-to-stellar mass relation}
  \label{sub:origin_gas_to_stellar}

  A gas-centric explanation treats the gas-to-stellar mass relation as the most fundamental of the three, with the SFMS following from the relations between molecular gas mass and stellar mass, and between molecular gas mass and SFR \citep{linALMaQUESTSurveyMolecular2019, feldmannLinkStarFormation2020, saintongeColdInterstellarMedium2022a}.
  This scenario leaves two questions open.
  First, it does not address which process determines the gas mass in the first place.
  Second, the gas fraction correlates strongly with galaxy size at fixed stellar mass \citep{zhangEstimatingHIGas2009, catinellaXGASSTotalCold2018}, and for disc galaxies their size is set by the angular momentum acquired from the host halo and the large-scale tidal torque field \citep{efstathiouStabilityMassesDisc1982, moFormationGalacticDiscs1998, shenSizeDistributionGalaxies2003, kravtsovSizeVirialRadiusRelation2013}.
  A gas-centric account is therefore complete only once it explains what sets the gas mass and where the gas fraction--size correlation comes from.

  We advocate instead that the gas fraction follows from the SFMS through the identity $M_\mg/M_\star = {\rm sSFR}/\epsilon$, with galaxy size linked to the star formation efficiency $\epsilon$.
  The link between galaxy size and star formation efficiency follows from the star formation law, although the direction of causality remains debated.
  One possibility starts from the Kennicutt star formation law, $\Sigma_{\rm SFR} \propto \Sigma_{\rm gas}^{1.4}$ \citep{kennicuttStarFormationGalaxies1998}, and suggests that galaxies with higher gas surface density have a higher star formation efficiency.
  Galaxies with a high surface density in gas build a stellar component of correspondingly high surface density and are therefore compact at fixed stellar mass.
  The other point of view states that the local stellar surface density regulates the star formation efficiency, $\Sigma_{\rm SFR} \propto \Sigma_\star^{0.4}\,\Sigma_{\rm gas}$ \citep{shiExtendedSchmidtLaw2011}.
  Either way, the size of a galaxy traces its global star formation efficiency.

  Our scenario therefore explains the correlation between gas fraction and galaxy size without further assumption: the identity $M_\mg/M_\star = {\rm sSFR}/\epsilon$ converts the variation in star formation efficiency associated with galaxy size into a variation in the gas fraction at fixed sSFR.
  The hierarchy further predicts that the correlation between galaxy size and SFR at fixed stellar mass is weak, since the star formation efficiency that size traces reaches the sSFR only through the transition between the inflow-driven and equilibrium regimes.
  This is what is observed \citep{wuytsGALAXYSTRUCTUREMODE2011, chuGalaxyStellarMassSFRSize2026}.

  \subsection{The origin of the mass--metallicity relation}
  \label{sub:mzr} 

  The framework in \citetalias{wanginflow1} shows that the gas metallicity is set by the gas fraction and the mass-loading factor, and the hierarchy established here adds that the gas fraction itself is the ratio of the sSFR to the star formation efficiency.
  The MZR is therefore shaped by three parameters: the halo formation time, the star formation efficiency, and the mass-loading factor, of which the first two enter through the gas fraction.
  These dependences produce the secondary correlations between gas metallicity and other galaxy properties at fixed stellar mass.

  At fixed stellar mass, galaxies with lower SFR, smaller size, or lower gas fraction are observed to have higher gas metallicity \citep{ellisonCluesOriginMassMetallicity2008, mannucciFundamentalRelationMass2010, yatesRelationMetallicityStellar2012, bothwellFundamentalRelationMetallicity2013, salimCriticalLookMassMetallicityStar2014a, jimmyGasPhaseMass2015, bothwellMolecularGasDriver2016, bothwellGalaxyMetallicitiesDepend2016, deugenioGasphaseMetallicitiesStarforming2018, curtiMassmetallicityFundamentalMetallicity2020, henryMassMetallicityRelation122021, maRevisitingFundamentalMetallicity2024, boardmanCompetingEffectsRecent2025, jiaPotentialdrivenMetalCycling2025, boardmanFundamentalMetallicityRelation2026}, and the same correlations are recovered in semi-analytic models \citep{yatesRelationMetallicityStellar2012, deluciaGasAccretionRegulates2020, wangOriginGalaxySizestellar2026} and in hydrodynamical simulations \citep{maOriginEvolutionGalaxy2016, derossiGalaxyMetallicityScaling2017, torreySimilarStarFormation2018, vanloonExplainingScatterGalaxy2021, linConstraintsGalacticOutflows2023, wangEnvironmentalDependenceMassMetallicity2023, garciaDoesFundamentalMetallicity2024, maRevisitingFundamentalMetallicity2024, garciaDoesFundamentalMetallicity2025, carnevaleMassDependenceFundamental2026}.
  Our framework accounts for these correlations.
  Galaxies with a lower gas fraction have a higher metallicity, and since the gas fraction is the ratio of the sSFR to the star formation efficiency, the anti-correlation with SFR follows immediately.

  The channel through which galaxy size modulates the metallicity is still under debate.
  One view treats size as an indicator of the depth of the gravitational potential, so that at fixed stellar mass compact galaxies have a deeper potential well, a lower mass-loading factor, and consequently a higher metallicity \citep{deugenioGasphaseMetallicitiesStarforming2018, sanchez-menguianoMoreFundamentalFundamental2024, sanchez-menguianoStellarMassNot2024, maRevisitingFundamentalMetallicity2024, jiaPotentialdrivenMetalCycling2025, kollerBothStellarMass2026}.
  A second view treats size as a tracer of the star formation efficiency \citep{wangOriginGalaxySizestellar2026} or the gas fraction \citep{boardmanFundamentalMetallicityRelation2026}.
  The hierarchy established here completes this chain: galaxy size traces the star formation efficiency, which modulates the gas fraction, and the gas fraction together with the mass-loading factor sets the gas metallicity.

  The two views can be distinguished cleanly.
  In our picture, the star formation efficiency reaches the metallicity only through the gas fraction, so controlling for the gas fraction should remove the size--metallicity correlation entirely.
  If instead size traces the mass-loading factor, which acts on the metallicity independently of the gas fraction, the correlation should persist at fixed gas fraction.
  Wang et al.\ (in preparation, hereafter Paper~III) carried out this test using the COLIBRE simulation \citep{schayeCOLIBREProjectCosmological2026, chaikinCOLIBRECalibratingSubgrid2026} and found that controlling for the gas fraction removes the correlation between galaxy size and gas metallicity, which disfavours the gravitational potential interpretation.

  A further view attributes the scatter of the MZR and its secondary correlations to stochastic fluctuations \citep{forbesOriginFundamentalMetallicity2014, torreySimilarStarFormation2018, deluciaGasAccretionRegulates2020, wangGasphaseMetallicityDiagnostic2021}.
  A sudden episode of gas inflow raises the gas mass, dilutes the metallicity, and enhances the SFR, and the subsequent decline of the inflow reverses all three.
  As noted in \citetalias{wanginflow1}, converting this temporal correlation within an individual galaxy history into a correlation across a population requires the galaxies to share a common mean, an assumption that is rarely made explicit and, to our knowledge, has not been justified.
  Paper-III applied a principal component analysis to COLIBRE central star-forming galaxies, separating the histories of gas fraction, sSFR, and gas metallicity into a long-timescale continuum ($\ga 3$~Gyr) and a short-timescale fluctuating part ($\la 3$~Gyr).
  The pairwise correlations among the three properties arise almost entirely from the long-timescale continuum.
  Combined with our earlier argument that the long-timescale variation of the SFR history cannot be stochastic without spoiling the SFMS, this indicates that the stochastic component plays a negligible role.

  \subsection{The role of halo mass}
  \label{sub:the_role_of_halo_mass} 

  \begin{figure}
  \begin{center}
    \includegraphics[width=0.95\linewidth]{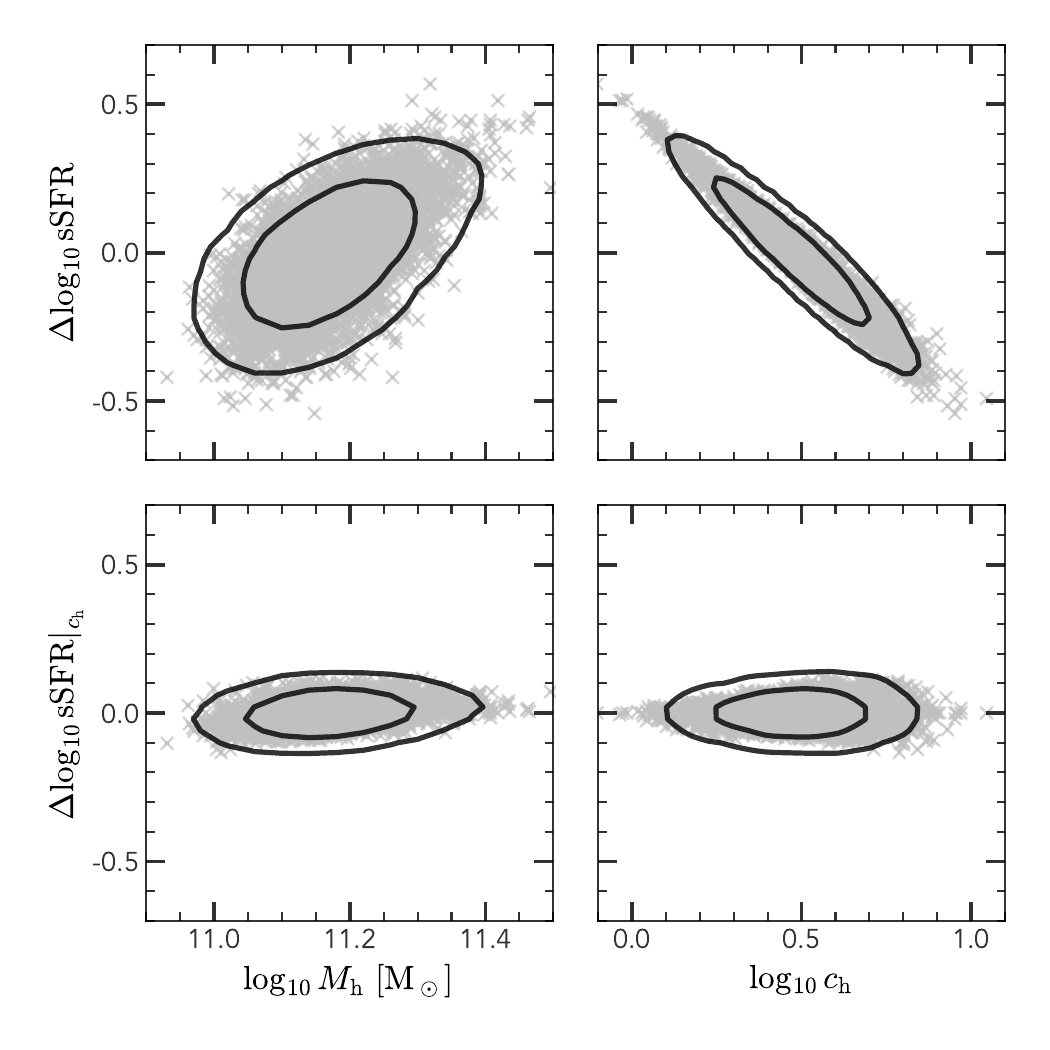}
  \end{center}
  \caption{
  The dependence of the sSFR on halo mass (left) and on the halo formation time parameter $c_\mh$ (right), for model galaxies (see \S\,\ref{sec:perspective_cosmological}) with $M_\star \approx 10^{9}\,\Msun$ at $z = 0$.
  The $y$-axis in the top row is $\Delta\log_{10}\mathrm{sSFR}$, the offset from the median sSFR at fixed stellar mass; in the bottom row it is the further residual after the dependence on $c_\mh$ has been subtracted in bins of width $0.1$~dex in $\log_{10} c_\mh$.
  Crosses show individual galaxies and the contours enclose 68 and 95 per cent of the population.
  The sSFR correlates with halo mass in the top left panel, but the correlation vanishes once the dependence on the formation time parameter is removed, whereas the anti-correlation with $c_\mh$ in the top right panel is much tighter.
  }
  \label{fig:role_mhalo}
  \end{figure}

  The halo mass is the primary property of a dark matter halo and plays an essential role in the galaxy-halo connection \citep{moGalaxyFormationEvolution2010, wechslerConnectionGalaxiesTheir2018}, so it is worth asking what part it plays in the three scaling relations.

  We first consider the ideal model in \S\,\ref{sec:perspective_ideal}.
  For a cooling efficiency $\lambda = 1$, the halo mass is $M_\mh = \Phi t / f_{\rm b}$, so it is related to the sSFR only through the shared dependence on the galaxy evolution time $t$.
  Fixing $t$ therefore removes the connection between halo mass and sSFR entirely.
  Neither the star formation efficiency nor the mass-loading factor depends on halo mass in the model, so the halo mass enters only by setting the stellar mass through the inflow rate $\Phi$.

  The minimal cosmological model in \S\,\ref{sec:perspective_cosmological} behaves the same way.
  Fig.~\ref{fig:role_mhalo} shows the joint distribution of the sSFR with halo mass and with the formation time parameter for galaxies of $M_\star \approx 10^{9}\,\Msun$.
  The top left panel shows a moderate positive correlation between the sSFR and halo mass, with galaxies in more massive haloes having higher sSFR at fixed stellar mass.
  The anti-correlation with $c_\mh$ in the top right panel is considerably tighter.
  The bottom row shows the residual sSFR after the dependence on $c_\mh$ has been subtracted in bins of width $0.1$~dex.
  The correlation with halo mass vanishes.
  This is what the ideal model anticipates: the apparent dependence on halo mass is inherited from the formation time, and controlling for the latter removes it.

  We caution that this conclusion is a property of the model, in which the star formation efficiency and the mass-loading factor are functions of stellar mass and redshift alone.
  Beyond the scope of this model, a dependence of the mass-loading factor on halo mass is physically plausible, since the outflow should escape the halo potential, and would introduce a halo mass dependence into the MZR.
  But the direction of that dependence is not obvious.
  \citet{wangGravitationalPotentialDrives2026} found, using the COLIBRE simulation, that at fixed stellar mass, the halo mass anti-correlates with the halo concentration, while both contribute positively to the depth of the potential, so which of the two dominates is not settled.
  Within this minimal framework, halo mass therefore plays no role in the three scaling relations beyond setting the stellar mass itself.
  Establishing whether it does so in reality will require a model in which the mass-loading factor is allowed to depend on it.

  \subsection{The chemical imprint on quiescent galaxies}
  \label{sub:quiescent}

  This work focuses on central star-forming galaxies, but the framework also bears on galaxies whose star formation has been quenched.
  Quenching may proceed through internal or external processes \citep{baldryGalaxyBimodalityStellar2006, pengMassEnvironmentDrivers2010, wooTwoConditionsGalaxy2015, whitakerPredictingQuiescenceDependence2017, wangDissectTwohaloGalactic2023}, and we restrict the discussion here to internally quenched centrals, since satellites quenched by environment have had their gas supply truncated from outside and their chemical evolution cannot be followed within our framework \citep{baheOriginEnhancedMetallicity2017, wangEnvironmentalDependenceMassMetallicity2023}.
  Such galaxies have negligible SFR and ISM mass, and no measurable gas metallicity, but their stellar populations retain a fossil record of their evolution.
  Here we consider the comparison of galaxy size and stellar metallicity between quiescent galaxies and their star-forming counterparts.

  At fixed stellar mass, quiescent galaxies have higher stellar metallicity than star-forming ones \citep{pengStrangulationPrimaryMechanism2015, lyuHalosGalaxiesVII2023, wangKinematicBimodalityEfficient2023}.
  \citet{vaughanSAMIGalaxySurvey2022} traced this difference to galaxy size: controlling for galaxy size removes the stellar metallicity difference entirely, so quiescence leaves no separate imprint on the stellar metallicity.
  Both ingredients of that explanation are established independently.
  Galaxy size anti-correlates at fixed stellar mass with the gas metallicity \citep{ellisonCluesOriginMassMetallicity2008, deugenioGasphaseMetallicitiesStarforming2018, sanchez-menguianoStellarMassNot2024, boardmanCompetingEffectsRecent2025, boardmanFundamentalMetallicityRelation2026} and with the stellar metallicity \citep{wangOriginGalaxySizestellar2026}, and quiescent galaxies have smaller galaxy size than star-forming ones at fixed stellar mass \citep{vanderwel3DHSTCANDELSEvolution2014, furlongSizeEvolutionNormal2017, genelSizeEvolutionStarforming2018, ludlowEvolutionSizesAngular2026}.

  Our framework supplies the mechanism.
  The stellar metallicity is the mass-weighted integral of the gas metallicity history, $Z_\star = (1-R)\int Z_\mg(t)\,{\rm SFR}(t)\,\dd t / M_\star$, so it inherits the correlations carried by the gas metallicity.
  In \S\,\ref{sub:mzr} and Paper-III we found that the anti-correlation between galaxy size and gas metallicity originates in the anti-correlation between galaxy size and star formation efficiency through the star formation law.
  The higher stellar metallicity of quiescent galaxies therefore indicates that their progenitors had smaller galaxy size, higher star formation efficiency, and consequently locked a larger fraction of their metals into the stellar component.

  Why, then, do the progenitors of quiescent galaxies have smaller galaxy size and higher star formation efficiency?
  One might invoke progenitor bias \citep{lillySURFACEDENSITYEFFECTS2016}, under which quenching selects from the star-forming population at random and the galaxy size difference merely reflects the redshift evolution of the mass--size relation \citep{vanderwel3DHSTCANDELSEvolution2014, ludlowEvolutionSizesAngular2026}.
  The stellar metallicity difference argues against this.
  If quiescent galaxies passively inherited the properties of high-redshift star-forming progenitors, they would have not only smaller galaxy size but also lower stellar metallicity, since the amplitude of the MZR decreases with increasing redshift \citep{maiolinoAMAZEEvolutionMassmetallicity2008, kashinoStellarMassStellar2022}.
  The observation runs the other way, quiescent galaxies being the more metal-rich population \citep{pengStrangulationPrimaryMechanism2015, lyuHalosGalaxiesVII2023}.
  Consistently, \citet{wangForgedQuenchingMorphological2026} carried out a controlled experiment in the EAGLE simulation \citep{schayeEAGLEProjectSimulating2015} and found that galaxy size plays an active role in quenching rather than being passively inherited through progenitor bias.

  Together with \citet{wangForgedQuenchingMorphological2026}, this gives a complete picture of the galaxy size and stellar metallicity difference between star-forming and quiescent galaxies.
  Galaxies form with a range of galaxy sizes, which are set by the angular momentum they acquire from their surroundings \citep{efstathiouStabilityMassesDisc1982, moGalaxyFormationEvolution2010}.
  Galaxies with smaller galaxy size have higher stellar surface density, and therefore higher star formation efficiency.
  This raises their gas metallicity, and through the integral above it also raises their stellar metallicity.
  These galaxies also host more massive black holes \citep{wangForgedQuenchingMorphological2026}, so AGN feedback in them is more powerful and they quench first.
  This is why the internally quenched population has smaller galaxy size and higher stellar metallicity.

  \section{Summary}
  \label{sec:summary}

  The star-forming main sequence (SFMS, $M_\star$--$\rm sSFR$), the gas-to-stellar mass relation ($M_\star$--$M_\mg/M_\star$), and the mass--metallicity relation (MZR, $M_\star$--$Z_\mg$) are the three core scaling relations of star-forming galaxies, and together they provide key insight into the baryon cycle.
  These three primary scaling relations are also correlated with each other through higher-dimensional scaling relations, such as the fundamental metallicity relation, gaseous fundamental metallicity relation, and the fundamental plane of star formation.
  These higher-dimensional correlations indicate strong connections among these three primary scaling relations.
  Here, we identify a hierarchy among these three scaling relations using both the analytic ideal gas flow model and the numerical minimal cosmological gas flow model.
  Our results are summarised as follows:
  \begin{enumerate}

  \item
  In the ideal gas flow model, in which the gas inflow rate, the star formation efficiency, and the mass-loading factor are held constant, the sSFR is set by the galaxy evolution time, with a weak modulation from the star formation efficiency and the mass-loading factor through the transition from the inflow-driven regime to equilibrium.
  The gas fraction follows from the sSFR and the star formation efficiency through the identity $M_\mg/M_\star = \mathrm{sSFR}/\epsilon$.
  The gas metallicity follows from the gas fraction and the mass-loading factor through the $\kk$ system, equations~\eqref{eq:master} and \eqref{eq:t2tau}.
  The three quantities therefore form a hierarchy: the sSFR is set primarily by the galaxy evolution time, the gas fraction by the sSFR together with the star formation efficiency, and the gas metallicity by the gas fraction together with the mass-loading factor (\S\,\ref{sec:perspective_ideal}).

  \item
  The same hierarchy holds in the minimal cosmological gas flow model in \citetalias{wanginflow1}, in which the gas inflow rate is modelled as the product of the halo-scale baryon accretion rate and the halo cooling efficiency, and the star formation efficiency and mass-loading factor are parameterised functions of stellar mass and redshift.
  The SFMS ($M_\star$--$\rm sSFR$) and its scatter are set primarily by the halo formation time and its scatter, with the star formation efficiency and the mass-loading factor playing a minor role through the transition between the inflow-driven regime and equilibrium.
  The gas-to-stellar mass relation ($M_\star$--$M_\mg/M_\star$) follows in turn from the SFMS together with the star formation efficiency and its scatter.
  The mass--metallicity relation ($M_\star$--$Z_\mg$) follows from the gas-to-stellar mass relation together with the mass-loading factor and its scatter (\S\,\ref{sec:perspective_cosmological}, Figs.~\ref{fig:cos_gas_const_offset}, \ref{fig:correlation_cos_gas_flow}, and \ref{fig:demo}).

  \item
  The hierarchy implies that scatter is inherited along the chain, but not that it grows at each step.
  The gas fraction scatters more than the SFMS, since the scatter of the star formation efficiency adds to that inherited from the sSFR, whereas the non-linear mapping onto the gas metallicity compresses the inherited scatter, making the MZR the tightest of the three relations.
  This ordering agrees with the observed scatters (Fig.~\ref{fig:sca_budget}).

  \item
  This hierarchy naturally introduces further correlations among the three relations, such as the fundamental plane of star formation ($M_\star$--$M_\mg$--$\mathrm{SFR}$), the fundamental metallicity relation ($M_\star$--$\mathrm{SFR}$--$Z_\mg$) and its gaseous counterpart ($M_\star$--$M_\mg$--$Z_\mg$) (Figs.~\ref{fig:obs_correlations} and \ref{fig:demo2}).

  \item
  The mass dependences of the star formation efficiency ($\epsilon \propto M_\star^{\epsilon_{\rm m}}$) and of the mass-loading factor ($\eta \propto M_\star^{\eta_{\rm m}}$) shape the slopes of the three relations.
  A larger $\epsilon_{\rm m}$ makes both the $M_\star$--$\mathrm{sSFR}$ and $M_\star$--$M_\mg/M_\star$ slopes more negative, and makes the MZR slope more positive.
  A larger $\eta_{\rm m}$ leaves the SFMS and the gas-to-stellar mass relation nearly unchanged, since the factor $1/(1-R+\eta)$ cancels in both the sSFR and $M_\mg/M_\star$, and makes the MZR slope less positive (see Fig.~\ref{fig:cos_gas_sfe_loading_m}).

  \item
  The redshift evolution of the sSFR at fixed stellar mass is set by the halo accretion history and is nearly independent of the star formation efficiency.
  As a $z=0$ halo is traced back to higher redshift, its specific accretion rate rises, so the same halo appears progressively later formed towards higher redshift.
  Late formation raises the sSFR, which sets the redshift evolution of the SFMS.
  The redshift dependence of the star formation efficiency ($\epsilon \propto (1+z)^{\epsilon_z}$) then governs the evolution of the gas fraction relative to that of the sSFR through the identity $M_\mg/M_\star = \mathrm{sSFR}/\epsilon$, and hence the evolution of the MZR: a larger $\epsilon_z$ lowers the gas fraction at high redshift and raises the metallicity at fixed stellar mass (see Fig.~\ref{fig:cos_gas_sfe_z}).

  \item
  The observed scatter and redshift evolution of the SFMS together bound the timescale on which stochastic variation could dominate the SFR history.
  The normalisation of the SFMS declines at $\approx 0.12$--$0.15$~dex~Gyr$^{-1}$ for $M_\star = 10^{9}$--$10^{10}\,\Msun$, while the scatter remains $\approx 0.2$~dex with no strong redshift dependence, so stochastic variations confined to that width cannot dominate on timescales longer than $\approx 1.3$--$1.7$~Gyr (\S\,\ref{sub:sfms}).

\item
Halo mass contributes negligibly to the three scaling relations beyond setting the stellar mass itself in our framework.
Its apparent correlation with the sSFR is inherited from the halo formation time and vanishes once the formation time is controlled for, and neither the star formation efficiency nor the mass-loading factor depends on halo mass in the model.
Settling whether halo mass plays a role beyond this will require more sophisticated models that can properly account for higher-order effects (see Fig.~\ref{fig:role_mhalo} and \S\,\ref{sub:the_role_of_halo_mass}).

\item
We extend the framework to explain why quiescent galaxies have higher stellar metallicity than star-forming ones at fixed stellar mass.
Quiescent galaxies have smaller galaxy size, hence higher star formation efficiency, higher gas metallicity throughout their history, and higher stellar metallicity in the end.
This disfavours progenitor bias, which predicts lower stellar metallicity for quiescent galaxies since their progenitors formed at high redshift (\S\,\ref{sub:quiescent}).

\end{enumerate}

This work identifies a hierarchy among the three core scaling relations of star-forming galaxies: the sSFR is set primarily by the halo formation time, the gas fraction follows from the sSFR together with the star formation efficiency, and the gas metallicity follows from the gas fraction together with the mass-loading factor.
The hierarchy accounts for the higher-dimensional relations reported in observations and in simulations, and for the secondary correlations of metallicity with SFR, size, and gas fraction at fixed stellar mass.
Reducing the model to the minimum within which the pattern can be identified analytically is what makes these dependences transparent.

\section*{Author contribution statement}
KW conceived the project, performed the analysis, led the interpretation of the results, produced all figures and wrote the manuscript.
All other co-authors, listed alphabetically, contributed to discussions of the results and commented on the manuscript.

\section*{Acknowledgements}
The authors thank the anonymous referee for their helpful comments that improved the quality of the manuscript.
KW thanks Peder Norberg, Andrew Pontzen, Yingjie Peng, Evgenii Chaikin, and Zheng Cai for stimulating discussions throughout this work.
KW acknowledges the use of Claude (Anthropic) as a writing and research aide in the preparation of this manuscript, including literature searches, drafting assistance, and language editing; all scientific content, analysis, and conclusions are the authors' own.

This work is supported by the Science and Technology Facilities Council (STFC) through grant ST/X001075/1.
KAO acknowledges support by the Royal Society through a Dorothy Hodgkin Fellowship (DHF/R1/231105).

This work used the DiRAC@Durham facility managed by the Institute for Computational Cosmology on behalf of the STFC DiRAC HPC Facility (www.dirac.ac.uk).
The equipment was funded by BEIS capital funding via STFC capital grants ST/K00042X/1, ST/P002293/1, ST/R002371/1 and ST/S002502/1, Durham University and STFC operations grant ST/R000832/1.
DiRAC is part of the National e-Infrastructure.

This research made use of NASA's Astrophysics Data System for bibliographic information.

\section*{Data Availability}
No new data were generated or analysed in support of this research. The model results can be reproduced from the equations and parameter values given in the text.



\bibliography{bibtex.bib} 
\bibliographystyle{mnras}



\appendix


\bsp  
\label{lastpage}
\end{document}